\documentclass[%
 aip,
 amsmath,amssymb,
reprint,%
]{revtex4-1}

\usepackage{graphicx}
\usepackage{dcolumn}
\usepackage{bm}

\usepackage[utf8]{inputenc}
\usepackage[T1]{fontenc}
\usepackage{mathptmx}
\usepackage{etoolbox}
\usepackage{enumitem}   
\usepackage{xcolor}
\makeatletter
\def\@email#1#2{%
 \endgroup
 \patchcmd{\titleblock@produce}
  {\frontmatter@RRAPformat}
  {\frontmatter@RRAPformat{\produce@RRAP{*#1\href{mailto:#2}{#2}}}\frontmatter@RRAPformat}
  {}{}
}%
\makeatother

\draft 

\begin{document}

\preprint{AIP/123-QED}


\title{Influence of the coordination defects on the dynamics and the potential energy landscape of two-dimensional silica}
\author{Projesh Kumar Roy}
\affiliation{The Institute of Mathematical Sciences, C.I.T. Campus, Taramani, Chennai 600113}
\affiliation{Homi Bhabha National Institute, Training School Complex, Anushakti Nagar, Mumbai 400094, India}
\affiliation{NRW Graduate School of Chemistry, Wilhelm-Klemm-Stra{\ss}e 10, 48149 M\"unster, Germany}

\author{Andreas Heuer}
\email{andheuer@uni-muenster.de}
\homepage{https://www.uni-muenster.de/Chemie.pc/en/forschung/heuer/index.html}
\affiliation{Institute f\"ur Physikalische Chemie, Westf\"alische-Wilhelms-Universitat M\"unster, Corrensstra{\ss}e 28/30, 48149 M\"unster, Germany}

\date{\today}


\begin{abstract}
The main cause of the fragile-to-strong crossover of 3D silica was previously attributed to the presence of a low-energy cutoff in the potential energy landscape. The important question emerges about the microscopic origin of this crossover and the generalizibility to other glass-formers. In this work, the fragile-to-strong crossover of a model 2D glassy system is analyzed via molecular dynamics simulation, which represents 2D-silica. By separating the sampled defect and defect-free inherent structures, we are able to identify their respective density of states distributions with respect to energy. A low energy cutoff is found in both distributions. It is shown that the fragile-to-strong crossover can be quantitatively related to the parameters of the energy landscape, involving in particular the low-energy cutoff of the energy distribution. It is also shown that the low-energy cutoff of the defect-states is determined by the formation energy of a specific defect configuration, involving two silicon and no oxygen defect. The low-temperature behavior of 2D silica is quantitatively compared with that of 3D silica, showing surprisingly similar behavior. 
\end{abstract}


\maketitle



\section{Introduction}

The origin of the anomalous temperature dependence of viscosity is a matter of debate for glassy systems. In general, the dynamical behavior of the glassy systems can be described either as `Strong' or `Fragile', depending on the nature of the relationship between the viscosity ($\eta$) and temperature (T); i.e. the Angell plot~\cite{AngellJNonCrysSolids1991} ($\ln(\eta(T))$ vs $\beta$, where $\beta=1/K_BT$). Assuming the validity of the Stokes-Einstein relation~\cite{Einstein_AnnPhysik_1906}, one can represent $\eta(T) \propto 1/D(T)$, where D(T) represents the diffusivity of the fluid. The strong glasses obey the Arrhenius-type temperature dependence of the diffusion constant, i.e. $D(T) \propto \exp(-\beta V_0)$, where $V_0$ is the average activation energy of the fluid. In contrast, the dynamics of the fragile glasses are usually described with the empirical Vogel–Fulcher–Tammann (VFT) equation~\cite{VogelPhysZ1921, FuchlerJAmCeramSoc1925,TammanZAnorgAllgChem1926, VFTJAmCerSoc1992}

\begin{equation}
D(T) \propto \exp \left ( \frac{\alpha}{T - T_0} \right )
\label{eqn:VFT}
\end{equation}

where $\alpha$, and $T_0$ are empirical constants. Experimentally, it has been shown that, several compounds undergo a Fragile-to-Strong crossover (FSC) at a certain temperature range~\cite{Angell_AnnRevPhysChem_1983, Angell_Science_1995, Hess_Rossler_ChemGeo_1996, Hemmati_Angell_JChemPhys_2001, Reibling_JChemPhys_1965, Lucas_Deymier_JPhysChemB_2017, Lucas_JNonCrysSolidsX_2019}, which is usually close to the glass transition temperature. Interestingly, molecules with tetrahedral ordering show a general tendency for such behavior in the bulk state~\cite{Shi_Tanaka_PNAS_2017, Russo_Tanaka_PNAS_2018, Shi_Tanaka_SciAdv_2019, Lucas_JNonCrysSolidsX_2019}. The fragile behavior of glasses at high temperatures--i.e. the power-law behavior of the relaxation time--can be explained with the help of mode-coupling theory~\cite{Marzio_Gallo_JChemPhys_2016, Starr_Stanley_PhysRevE_1999}. Upon cooling, the transition to Arrhenius behavior is often correlated with a supposed liquid-liquid phase transition, which generally occurs in the super-cooled region~\cite{Xu_PNAS_2005, Poole_Stanley_Nature_1992, Mishima_Stanley_Nature_1998, Palmer_Debenedetti_Nature_2014,Poole_Stanley_PhysRevE_1993,Poole_Stanley_PhysRevE_1993_2}. However, a proper physical mechanism for FSC--at either bulk or molecular level--is not well understood. Potential energy landscape (PEL) analysis of the glassy systems provides a good understanding of the FSC phenomenon at low temperatures. It is known that, below the mode-coupling temperature, the dynamics and the thermodynamics of the system start to correlate, and the nature of the dynamical properties strongly reflect the properties of the PEL. In this region, the nature of the PEL can be described by the properties of the underlying minima, i.e. Inherent States (IS)~\cite{Stillinger_Weber_PhysRevA_1982, Stillinger_Weber_PhysRevA_1983, Stillinger_Weber_Science_1984, Stillinger_Weber_JPhysChem_1983, Weber_Stillinger_JChemPhys_1984, Sastry_Stillinger_Nature_1998, Debnedetti_Stillinger_Nature_2001, Heuer_JPhysCondMatt_2008}. At higher temperatures, the fragile behavior can be linked to the multiplicity and the width of the underlying IS-energy distributions~\cite{Sastry_Nature_2001, Martinez_Angell_Nature_2001}. With decreasing temperature, the configurational entropy of the system ($S_c$) decreases, and the widths of the basins in the PEL landscape become narrower. Following the Adam-Gibbs theory~\cite{Adam_Gibbs_JChemPhys_1965}, one can determine the relaxation time and subsequently the diffusivity as a function of the $S_c$ in the low-temperature region as $D(T) \propto \exp(-B/TS_c)$, where B is a constant. Therefore, if $S_c$ is independent of temperature~\cite{Kaori_Angell_Nature_1999, Voivod_Sciortino_Nature_2001, Voivod_Poole_PhysRevE_2004}, a transition from fragile to strong behavior can be related to an inflection in $S_c$. Using molecular dynamics simulations, Saksaengwijit et al.~\cite{HeuerPhysRevLett2004} had shown that in case of BKS-silica~\cite{BeestPhysRevLett1955}, there exists a cutoff in the density of states (DOS) distribution of IS's at a certain energy. The presence of such a cutoff prevents $S_c$ from vanishing at low temperatures and thus avoids the Kauzmann paradox. However, determination of the nature of reduction of $S_c$ at low temperature is a difficult task in simulation as it is computationally demanding~\cite{Yu_Wang_PhysRevMater_2021, Yu_Wang_arxiv_2022}. Thus, proper calculation of the FSC temperature from the IS analysis remains challenging.

The microscopic origin of FSC varies from system to system. Poly-amorphism~\cite{Voivod_Poole_PhysRevE_2004} was often suspected to be the drive behind the FSC phenomenon. For example, a two-state model was proposed by Tanaka et al.~\cite{Tanaka_JPhysCondMatt_2003, Shi_Tanaka_PNAS_2018, Shi_Tanaka_JChemPhys_2018} to explain the viscosity anomaly of water. In this model, two different configurations of \textit{locally favored} small clusters of water molecules controls the bulk properties of water. The FSC in water can thus be explained from the population ratio of such clusters, both of which separately follow Arrhenius type temperature dependence of viscosity. This model predicts a smooth crossover from the fragile to the strong regime, and was able to predict the crossover temperature, $T_{\text{FSC}}$, quite well~\cite{Shi_Tanaka_PNAS_2018, Shi_Tanaka_JChemPhys_2018}. In case of several tetrahedrally ordered halide-molecules~\cite{Lucas_JNonCrysSolidsX_2019}; e.g. ZnCl$_2$~\cite{Zeidler_Howells_JNonCrystSolids_2015, Lucas_Deymier_JPhysChemB_2017}, BeF$_2$~\cite{Hemmati_Angell_JChemPhys_2001, Zeidler_Howells_JNonCrystSolids_2015}, GeSe$_2$~\cite{Petri_Fischer_PhysRevLett_2000, Crichton_Grzechnik_Nature_2001, Edwards_Sen_JPhysChemB_2011}; the reduction of the $S_c$ near FSC has also been correlated to the changes in the ordering of the local environment. For these materials, in both experiment~\cite{Edwards_Sen_JPhysChemB_2011, Zeidler_Howells_JNonCrystSolids_2015, Lucas_Deymier_JPhysChemB_2017} and simulation~\cite{Gupta_Mauro_JChemPhys_2009, Wilson_Salmon_PhysRevLett_2009}, a sharp change has been observed in the ratio of the local edge-sharing tetrahedral ordering to entropically favored corner-sharing tetrahedral ordering, with lowering temperature in the vicinity of FSC. These results clearly indicate that FSC is linked to the specific microscopic environment of the fluid. Recently a continuum model, in conjugation to the two-state theory, is proposed to explain the non-Arrhenius behavior of the liquids using non-linear Fokker-Plank equations~\cite{Rosa_Moret_PhysRevE_2019, Rosa_Moret_PhysRevE_2020}.

Despite the similarity between the molecular structure (tetrahedral) of water and silica, they exhibit several differences in terms of cluster formation~\cite{Shi_Tanaka_PNAS_2018_2}. While the structure of the water-clusters are controlled by the hydrogen-bond network, silica-clusters are controlled by the flexibility of the bridging-oxygen bonds. As a result, although water and silica exhibits some dynamical and thermodynamical similarity--e.g. density anomaly, FSC, etc.--their microscopic origin can be somewhat different. Saksaengwijit et al.~\cite{HeuerPhysRevLett2004} had postulated that for BKS-silica~\cite{BeestPhysRevLett1955}, the FSC can be related to a decrease in the fraction of coordination defects. As the temperature of the system is lowered, the fraction of the coordination defects in silica decreases. Recently, using replica exchange molecular dynamics, Yu et al.~\cite{Yu_Wang_PhysRevMater_2021} have shown that the rate of such a decrease shows a sharp increase near its fictive temperature. Thus, near the cutoff, one can postulate that the defect states would completely disappear. This, in turn, should result in an arrest in the hopping dynamics of silica as the effective activation energy for the bond-breaking process would be very large~\cite{HeuerPhysRevLett2004, Heuer_PhysRevE_2006}.

In this paper, we study the FSC behavior of a 2D model system which represents an allotrope of silica, namely the two-dimensional silica~\cite{HeydeAngewChem2012} (2D-silica). This material is an ideal member to study the properties of a natural glass forming system with computer simulation since an atomistic scale description of the real structure is available directly from the experiment. 2D-silica is grown on Ru~\cite{HeydeJPhysChemC2012} or Graphene surfaces~\cite{Huang_Kaiser_NanoLett_2012} via vapor deposition and in-situ oxidation. The binding of 2D-silica with the underlying surface is quite weak which makes it a free-standing material~\cite{HeydeChemPhysLett2012}. However, the material is extremely thin with a height of just $\sim$ 3 atoms in the Z-direction~\cite{HeydeJPhysChemC2012}. Therefore, measuring high temperature properties of this material is quite difficult as it would desorb from the surface during heating. Thus, for this system the FSC behavior is expected to be inaccessible via experiments. Therefore, computer simulations play an important role to elucidate the properties of this system close to the FSC.

The objective of this paper is to show that the PEL description of the dynamical behavior works well for the description of FSC in the 2D-silica model. Similar to the BKS-silica model, we aim to correlate the presence of a cutoff in the PEL with the FSC phenomenon in 2D-silica. We provide further insight into the microscopic environment of 2D-silica near the FSC in terms of the coordination defects. The paper is organized as follows. We start with a short account of the methods in section \ref{sect:methods} and analyze the density of states (DOS) distributions. In section \ref{sect:PEL}, we study the changes in the PEL influenced by the coordination defects and explicitly show that the FSC is related to the cutoff in the PEL. Afterwards, we show in section \ref{sect:structure} that a specific local arrangement of the particles is present at the low energy limit which is independent of the system size, thus yielding insight into the structural and energetic properties of the defects. We finish with a conclusion and outlook.


\section{Methods}
\label{sect:methods}

\subsection{Simulation}

In our previous works~\cite{Roy_Heuer_PCCP_2018, Roy_Heuer_PRL_2019, Roy_Heuer_JPhysCondMatt_2019}, we have used a Yukawa type force-field to describe the structural and thermodynamical properties of 2D-silica in close similarity with the experimental results, which reads

\begin{equation}
  V_{ij}(r_{ij}) = \left [\left ( \frac{\sigma_{ij}}{r_{ij}} \right )^{12} + \left ( \frac{q_{ij}}{r_{ij}} \right ) \exp(-\kappa r_{ij}) \right ]
  \label{eqn:yukawa}
\end{equation}

For details of the simulation parameters, we refer to the reference~\citenum{Roy_Heuer_PCCP_2018}. With a 80 particle system in a square box of side length 19.67 {\AA}, we performed a molecular dynamics simulation under NVT conditions for a wide range of temperatures using Nos\'{e}-Hoover chains thermostat~\cite{NoseJChemPhys1984, MartynaJChemPhys1992}, and an elementary timestep of 0.01. We sample the inherent structures and their energies ($E_{\text{IS}}$) by quenching the trajectory with a frequency of 100 steps. Except for the mean square displacement (MSD) calculations, we use the minimized structures for analysis throughout the paper.

\subsection{Defects}

In our analysis, we separate the IS's based on the coordination defects. We denote an IS as a defect-free-state (abbreviation: df), if all Si particles have 3 O particles and all O particles have 2 Si particles within an empirical cutoff distance of 2.0 {\AA}. Otherwise the IS is termed as defect-state (abbreviation: d). The choice of this cutoff is based on the g$_{\text{Si-O}}$(r) analysis~\cite{Roy_Heuer_PCCP_2018}, which showed a sharp peak around 1.5-1.6 {\AA} at the low energy regime. Since we are interested in the structural features of the model at low temperatures, the choice of the distance cutoff is sufficient in identifying any coordination defects in the structure. 

\subsection{Reweighting}

With the sampled IS structures at one given temperature, one can calculate the DOS ($G(E_{\text{IS}})$) at different energies. This calculation is performed by binning the distribution of inherent structures $P(E_{\text{IS}}, T)$ and reweighting the resulting energy distribution with an inverse Boltzmann factor, i.e.

\begin{equation}
G(E_{\text{IS}}) \propto P(E_{\text{IS}}, T)  e^{\beta E_{\text{IS}}}
\label{eqn:DOS}
\end{equation}

Naturally, one unknown energy-independent proportionality constant remains. Similarly, one can define $G_{\text{d}}(E_{\text{IS}})$ and $G_{\text{df}}(E_{\text{IS}})$ as the DOS of the defect-states and defect-free-states, respectively. We note that this relation, i.e. the estimation of a temperature independent DOS, is valid as long as anharmonic effects in the vibrations within an IS are not too strong~\cite{HeuerPhysRevE1999, Heuer_JPhysCondMatt_2008}. Strictly speaking the DOS should be rather interpreted as an effective DOS because each state is also weighted by the harmonic partition function, i.e by the inverse square root of the frequency \cite{Heuer_JPhysCondMatt_2008}. It is known that the DOS distribution of a binary Lennard-Jones liquid or BKS-silica has a Gaussian shape over a large energy range~\cite{HeuerPhysRevE1999, HeuerJPhysCondMatt2000, HeuerPhysRevE2008}.

To determine $G(E_{\text{IS}})$ over a broad energy range, based on the Boltzmann distributions from different temperatures, one could try to overlap the different $G(E_{\text{IS}})$-curves, obtained from different temperatures. A procedure to obtain the DOS in a statistically optimized way is the Weighted Histogram Analysis Method (WHAM)~\cite{Kumar_Kollman_JCompChem_1992}. To obtain the DOS $G(E_{\text{IS}})$ we use the algorithm developed by Gallicchio {\it et al.}~\cite{GallicchioJPhysChemB2005} which reads for the present application

\begin{equation}
G(E_{\text{IS}}) = \frac{ \sum_i N(E_{\text{IS}}, T_i) }{ \sum_i N_i(T_i) f_i(T_i) c_i(E_{\text{IS}}, T_i) }
 \label{eqn:WHAM}
\end{equation}

Here, $N(E_{\text{IS}}, T_i)$ is the total counts of the states at energy bin $E_{\text{IS}}$ at temperature $T_i$ and $N(T_i) = \sum_{j} N(E^j_{\text{IS}}, T_i)$. $c_i(E_{\text{IS}}, T_i)$ is called a component of the {\it biasing matrix}, which is calculated as

$$c_i(E_{\text{IS}}, T_i) = \exp [ -\beta_i E_{\text{IS}} ]$$

From there, one can optimize the normalization factor $f_i(T_i)$ as

$$ f_i^{-1}(T_i) = \sum_j c_i(E^j_{\text{IS}}, T_i) G(E^j_{\text{IS}}, T_i)$$

We iteratively optimize the normalization factor $f_i(T_i)$ until convergence is reached and the DOS can be read off. Note that the resulting DOS is defined except for one proportionality constant.

In analogy this procedure can be applied as well for the determination of $G_{\text{d}}(E_{\text{IS}})$ and $G_{\text{df}}(E_{\text{IS}})$. However, due to the unknown proportionality constants we would not be able to determine, e.g., the fraction of defect-free-states $q_{\text{df}}(E_{\text{IS}})$ at a given energy $E_{\text{IS}}$. Naturally, from equilibrium simulations at a given temperature $T_i$ one can directly read of this fraction $q_{\text{df}}^i(E_{\text{IS}}, T_i)$. Since this fraction is an equilibrium property, in the limit of infinite sampling it would not depend on temperature. In the spirit of WHAM we join the information from the simulations at the different temperatures for the estimation of $q_{\text{df}}(E_{\text{IS}})$ via
\begin{equation}
 q_{\text{df}}(E_{\text{IS}}) = \frac{ \sum_i N(E_{\text{IS}}, T_i) q_{\text{df}}^i(E_{\text{IS}}, T_i) }{ \sum_i N(E_{\text{IS}}, T_i) }
 \label{eqn:average_fraction}
\end{equation}

In practice we first determine $G(E_{\text{IS}})$ (via WHAM) and $q_{\text{df}}(E_{\text{IS}})$ as described above and then estimate the DOS for the defect-free-states and defect-states, respectively, via
\begin{eqnarray}
 G_{\text{df}}(E_{\text{IS}}) = G(E_{\text{IS}}) q_{\text{df}}(E_{\text{IS}}) \nonumber \\
 G_{\text{d}}(E_{\text{IS}}) = G(E_{\text{IS}}) (1 - q_{\text{df}}(E_{\text{IS}}))
 \label{eqn:DOS_WHAM}
\end{eqnarray}


\section{PEL properties and their relation to the diffusivity}
\label{sect:PEL}

In this section, we investigate the DOS distributions of defect and defect-free states with energy using the WHAM optimized energy distributions in the previous section. All important parameters, to be discussed below, are summarized in Table~\ref{tbl:params}. 

\begin{table}[h]
	\caption{Values of the important variables used in section~\ref{sect:PEL}. For the description of the variables, see main text and supporting material (SM).}
\begin{tabular}{|c|c|}
	\hline
	$E^{\text{cutoff}}_{\text{df}}$ (Figure~\ref{fig:DOS_defect_defectfree}(c)) & -27.04 \\
	\hline
	$E^{\text{cutoff}}_{\text{d}}$ (Figure~\ref{fig:DOS_defect_defectfree}(c))  & -26.93 \\
	\hline
	$E^{\text{cross}}_0$ (Figure~\ref{fig:DOS_defect_defectfree}(c))  & -26.86 \\
	\hline
	$E^{\text{max}}_{\text{IS}}$ (Figure SM.1(a))  & -25.77 \\	
	\hline
	$E^{\text{max}}_{\text{d}}$ (Figure~\ref{fig:DOS_defect_defectfree}(b))  & -25.89 \\	
	\hline
	$E^{\text{max}}_{\text{df}}$ (Figure~\ref{fig:DOS_defect_defectfree}(c))  & -26.56 \\	
	\hline
	$s_{\text{IS}}$ (Figure SM.1(a))	&   0.12	  \\
	\hline
	$s_{\text{d}}$ (Figure~\ref{fig:DOS_defect_defectfree}(b))	 &   0.12	  \\
	\hline
	$s_{\text{df}}$ (Figure~\ref{fig:DOS_defect_defectfree}(c))	 &   0.07	  \\
	\hline
	$V_0$ (Figure~\ref{fig:Diffusion_constant}(a))  &   0.26	  \\
	\hline
\end{tabular}
\label{tbl:params}
\end{table}

\subsection{Density of states}
\label{subsect:density}

\begin{figure*}[!htb]
 \includegraphics[scale=1.0]{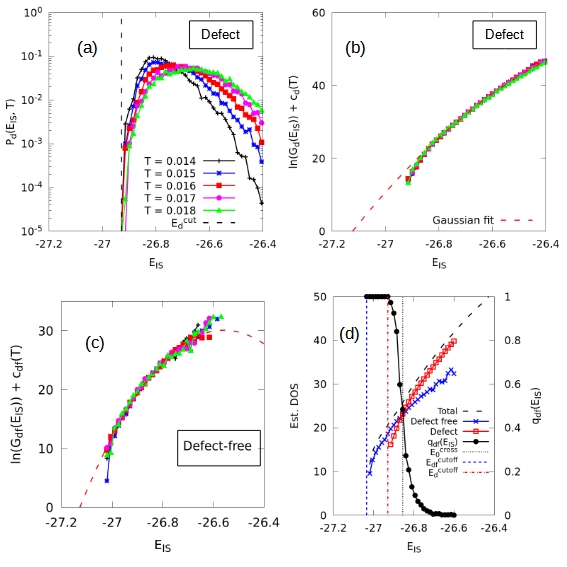}
 \caption{(a){ The semi-log plot for the normalized Boltzmann distributions of the defect-states for temperatures between T=0.014 and T=0.018. One can estimate the cutoff energies for the defect-states at $\sim$ -26.93.} (b) DOS of defect-states, obtained from the data in (a) via Boltzmann reweighting. The data at T=0.015 are fitted with a Gaussian function $\ln G(E) = -\frac{1}{2}[(E - E^{\text{max}}_{\text{d}})/s]^2 + c_{\text{d}}(T)$ with $E^{\text{max}}_{\text{d}} = -25.89, s_{\text{d}}=0.12$ and an arbitrary shift parameters $c_{\text{d}}(T)$. (c) DOS of defect-free states, obtained is a similar way as (b) with $E^{\text{max}}_{\text{df}} = -26.56, s_{\text{df}}=0.07$. (d) The plot for the estimated DOS from the WHAM analysis for defect-states, defect-free-states, and all states. The total DOS curve (marked as \textit{Total}) is slightly shifted along the Y-axis for better visual inspection. The fraction of defect-free states at each energy level ($q_{\text{df}}(E_{\text{IS}})$) is also shown and its values are related to the right Y-axis. For a complete list of different parameters, see Table~\ref{tbl:params}.}
 \label{fig:DOS_defect_defectfree}
\end{figure*}

In Figure~\ref{fig:DOS_defect_defectfree}(a) we show the Boltzmann distribution $p_{\text{d}}(E_{\text{IS}}, T)$ of defect-states for a large range of temperatures. Whereas for the highest temperature (curve at the right end) the distribution is basically Gaussian, one observes for lower temperatures deviations from the Gaussian shape in the low-energy range and the presence of a cutoff-energy of $E_{\text{d}}^{\text{cutoff}} = -26.93$. This value does not seem to depend on temperature.

In the next step, we reconstructed the underlying DOS via simple reweighting (see Figure~\ref{fig:DOS_defect_defectfree}(b)). Indeed, we find that over a large range of energies DOS curves overlap with each other, which confirmed that the simulations at the lowest temperatures have been done at equilibrium. Similar observation can be made for the total DOS and the defect-free DOS as well (see supporting material, Figure SM.1). In the high energy range, the DOS seems to depend on temperature. As mentioned above, this indicates the presence of significant anharmonic effects for high temperatures. For a large range of energies the distribution follows a Gaussian shape. Importantly, in the low-energy range the DOS of defect-states decays much faster than a Gaussian distribution, reflecting the presence of the cutoff-energy. The analogous results for the defect-free-states are shown in Figure~\ref{fig:DOS_defect_defectfree}(c) (see also supporting material Figure SM.1(a), for DOS of all-states). Finally, in Figure~\ref{fig:DOS_defect_defectfree}(d) we show the results of the WHAM analysis, i.e. the DOS $G(E_{\text{IS}})$, $G_{\text{d}}{\text{IS}})$, $G_{\text{df}}(E_{\text{IS}})$, as well as the fraction of defect-free-states $q_{\text{df}}(E_{\text{IS}})$. Note that the relevant energy regime, shown in this figure, only covers the regime where the determination of the DOS is not hampered by anharmonic effects. Also for the defect-free-states we find a low-energy cutoff. The estimated cutoff energy of the defect-free-states, $E^{\text{cutoff}}_{\text{df}} \approx -27.04$ is significantly lower than $E^{\text{cutoff}}_{\text{d}}$, namely $\Delta E^{\text{cutoff}} = E_{\text{d}}^{\text{cutoff}} - E_{\text{df}}^{\text{cutoff}} \sim 0.11$. As a consequence, also the overall DOS $G(E_{\text{IS}})$ displays the same cutoff as $G_{\text{d}}(E_{\text{IS}})$. Such a cutoff-energy was also observed in case of BKS-silica~\cite{HeuerPhysRevLett2004}.

Interestingly, $\Delta E^{\text{cutoff}}$  hardly changes when increasing the system size from 80 to 200 particles ($\Delta E^{\text{cutoff}}(N=200) \sim 0.12$ as can be derived from the data in supporting material Figure SM.2). This result suggests that the minimal number of particles, required to create a defect state, is independent of size. Later we will show that the structure of the defect-states at the cutoff energy involves a well-defined local environment, other than the usual trigonal geometry so that the energy of 0.11 can be interpreted as a minimum defect formation energy. Furthermore, $E_{\text{df}}^{\text{cutoff}}/N$ hardly varies with system size. {Thus the energy of the lowest energy state, which is a defect-free state, is an extensive property. Due to the local character of the formation of a defect it is expected that $\Delta E^{\text{cutoff}}$ is not extensive.} This is indeed clearly seen in supporting material Figure SM.2.

Finally, we discuss the properties of the fraction of defect-free-states $q_{\text{df}}(E_{\text{IS}})$. It has an asymmetric behavior: whereas at low energies one observes a sharp cutoff, below which only defect-free-states are available, at high temperatures there is a more gradual approach towards the limit where only defect-states are present. Indeed, as seen from the supporting material (Figure SM.1), there is no indication of a strict cutoff-energy at high energies, although defect-free-states become exponentially rare. Nevertheless, there is a fast transition from the energy regime governed by defect-free-states to the regime governed by defect-states. In fact, it takes an energy variation of less than 0.06 to change between $q_{\text{df}}(E_{\text{IS}})=0.2$  and $q_{\text{df}}(E_{\text{IS}})=0.8$.  The crossover energy, $E^{\text{cross}}_0$, where $q_{\text{df}}(E_{\text{IS}})=0.5$, i.e. where the number of defect-free-states and defect-states is identical, turns out to be -26.86. In terms of per-particle energies, the value of $E^{\text{cross}}_0$ gets lower for a larger system size, see the supporting material (Figure SM.2). This is because the probability of finding defect-free-states decreases exponentially with increasing system size~\cite{Roy_Heuer_PCCP_2018}. This is a consequence of the fact that a large system can be decomposed into approximately independent smaller subsystems. Then it becomes very unlikely that the total system contains no defect because this requires that every subsystem is defect-free. This underlines the importance of studying a smaller system sizes where the statistical characterization of defect-free-states is far easier.

\subsection{Fragile to strong crossover}
\label{subsect:FSC}

\begin{figure*}[!htb]
\centering
 \includegraphics[scale=1.0]{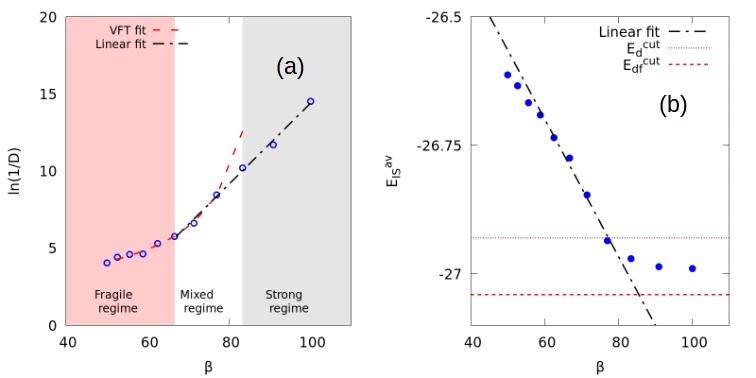}
 \caption{{(a) Dependence of $\ln(1/D)$ on inverse temperature. The data in the fragile-regime (region marked in red) are fitted with the VFT equation (equation~\ref{eqn:VFT}) and data in the strong-regime (region marked in gray) are fitted with an Arrhenius equation (equation~\ref{eqn:diffusion}). Values of $\alpha$, and $T_0$ are $\approx$ 0.014, and 0.011, respectively. The value of $V_0$ is $\approx$ 0.26. Data inside the mixed regime are not fitted using any equations, but the approximations from both fragile and strong regime are extrapolated to measure $\beta_{FSC}$. (b) Dependence of the average IS energy on the inverse temperature. A linear fit is performed between $62.5 \leq \beta \leq 77$ with slope $\sim -0.013$. The cutoff energy for the total system (same as for the defect-free-states, $E_{\text{df}}^{\text{cut}}$) and defect states ($E_{\text{d}}^{\text{cut}}$) are also shown as horizontal lines.}}
 \label{fig:Diffusion_constant}
\end{figure*}

Next we explore a possible relation between the kinetics and the thermodynamics of the system. In Figure~\ref{fig:Diffusion_constant}(a) we show the temperature-dependent diffusion constant. One clearly finds a transition from non-Arrhenius to simple Arrhenius behavior,

\begin{equation}
D(T) = D_0 \exp [-V_0 \beta],
\label{eqn:diffusion}
\end{equation}
i.e. a transition from fragile to strong behavior. We have added fits with the VFT equation in the high-temperature and an Arrhenius fit in the low-temperature regime. Starting from the VFT fit, the first significant deviation is seen for $\beta \approx 80$.

To relate the kinetics to the thermodynamics of the system, we provide the average IS energy ($E_{\text{IS}}^{\text{av}}(T)$ in Figure~\ref{fig:Diffusion_constant}(b). Again, two temperature regimes can be observed. Both have a simple interpretation. If the DOS is governed by a Gaussian distribution with variance $s^2$ one expects a linear behavior
\begin{equation}
E_{\text{IS}}^{\text{av}}(T) = E_{\text{IS}}^{\text{max}} - \beta s^2.
\label{eqn:IS_avE}
\end{equation}
Indeed, this is seen in the high-temperature regime in Figure~\ref{fig:Diffusion_constant}(b). The negative slope of 0.013 is very close to the result of the Gaussian fit ($s^2 = 0.014$), reported in Table~\ref{tbl:params}. Naturally, in the low-temperature regime the average energy starts to approach the cutoff-energy. Most importantly, the impact of the cutoff-energy again starts to be visible around $\beta \approx 80$ which is exactly the same temperature regime as observed for the FSC. This strongly suggests a direct causal relation between the kinetics and the thermodynamics. We mention in passing that the crossover temperature can be estimated from the condition $E_{\text{IS}}^{\text{av}}(T) \approx E^{\text{cutoff}}_{\text{df}}+ s$. The second summand expresses the fact that the energy-cutoff starts to matter as soon as the cutoff is within a standard deviation to the average energy (in Gaussian approximation). Using the values $s = 0.12, E_{\text{IS}}^{\text{max}}= -25.77$ (see supporting material, Figure SM.1), and $E^{\text{cutoff}}_{\text{df}}= -27.04$, one obtains $\beta = 79.8$ which is basically identical to the estimated crossover temperature of $\beta \approx 80$.

\subsection{Comparison with bulk silica}

{Previously, FSC has been observed also for 3D bulk silica--simulated with the standard BKS forcefield~\cite{HeuerPhysRevLett2004}--and related to the properties of the energy landscape. In this way, the causality could be directly proven~\cite{Heuer_PhysRevE_2006}. Here we perform a quantitative comparison of the FSC in order to identify similarities and differences between 2D silica and 3D silica.} However, before starting we estimate the crossover temperature for 3D ($s = 3.5$ eV, $E_{\text{IS}}^{\text{max}}= -1867$ eV, and $E^{\text{cutoff}}_{\text{df}}= -1910.7$ eV) as $T = 0.30$ eV $\approx 3500$ K which agrees very well with the emergence of the crossover behavior of the diffusivity and the average energy, reported in reference~\citenum{HeuerPhysRevLett2004}.

\begin{enumerate}[label=(\roman*)]

\item \textit{System size:} For the estimation of the diffusivity one can define an energy-dependent activation energy $E_{\text{eff}}(E_{\text{IS}})$~\cite{Heuer_PhysRevE_2006} which for 3D silica can be written to a very good approximation as (for $E_{\text{IS}} \ge E^{\text{cutoff}}_{\text{df}}$)

\begin{equation}
E_{\text{eff}}(E_{\text{IS}}) \approx \lambda (E^{\text{cutoff}}_{\text{df}} - E_{\text{IS}})+  V_0(E^{\text{cutoff}}_{\text{df}}).
\label{eqn:activation}
\end{equation}

More generally, the possibility to formulate a relation between activation energy and IS energy (strictly speaking metabasin energy, which, however, for the typically visited IS at low temperatures is basically identical) shows the generally strong connection between the dynamic properties and the nature of the PEL for glass-forming systems~\cite{Sastry_Stillinger_Nature_1998, DoliwaPhysRevLett2003, DoliwaPhysRevE2003, HeuerPhysRevE1999}. The value $\lambda \le 1$ has a simple interpretation: $\lambda \times N$ is the size of the hypothetical elementary subsystem to which relaxation processes are basically confined. Indeed it has been shown~\cite{Heuer_JPhysCondMatt_2008} that by doubling the system size the value of $\lambda$ decreases by a factor of 2. From the data in reference ~\citenum{Heuer_PhysRevE_2006} one can estimate $\lambda \approx 0.6$. Thus, for the system size of $N=99$, used for the simulation of bulk silica, an elementary system comprises approx. 60 particles.
 
For the present case we have no direct information for the estimation of $\lambda$. However, we can formulate an upper bound. Naturally, for the highest temperature where anharmonic effects do not matter $(\beta \approx 60)$ one still has activated behavior, i.e. $E_{\text{eff}}(E_{\text{IS}}^{\text{av}}(\beta = 60)) \gtrsim 0$. This implies (using $V_0(E^{\text{cutoff}}_{\text{df}})=0.26$, see Figure~\ref{fig:Diffusion_constant}(a), and $E_{\text{IS}}^{\text{av}}(\beta = 60)= -26.69$) that $\lambda \le 0.74$. Thus, the hypothetical elementary subsystem has a size of less than $0.74 \times 80 \approx 60$ particles.

\item \textit{PEL landscape:} A natural dimensionless number, characterizing the shape of the DOS, is given by $ [E_{\text{IS}}^{\text{max}} - E^{\text{cutoff}}_{\text{df}})]/s$. It expresses the impact of the cutoff-energy relative to the overall shape of the Gaussian. In the present case we have (-25.77 + 27.04)/0.12 = 10.6. Reading off the corresponding values for bulk silica from reference \citenum{HeuerPhysRevLett2004}, we obtain (-1867 eV + 1910.7 eV)/3.5 eV = 12.5. However, for a direct comparison we need to take into account that these ratios depend on system size. Whereas the energies in the numerator scale with the system size, the standard deviation in the denominator scales with the square root of the system size. For a fair comparison the values should be evaluated for the respective elementary system sizes. However, that with the present data it cannot be excluded that the elementary system sizes are comparable, the same holds for the shape of the DOS.  

\item \textit{Defect-states vs. defect-free-states:} One may ask for the energy to generate defects. In the present case it is given by $\Delta E^{\text{cutoff}}\approx 0.11$. To obtain an appropriate dimensionless number we may relate this value to the crossover temperature, yielding $(0.11/(1/80)) \approx 8.8$. Interestingly, for bulk silica there exist defect-states very close to the cutoff-energy of defect-free-states. From the data in reference~\citenum{HeuerPhysRevLett2004} one may estimate $\Delta E^{\text{cutoff}}_{\text{3D}}\approx 1.0$ eV. This yields a ratio of (1 eV/0.3 eV) = 3.3 which is significantly smaller than 8.8. One may speculate that in 3D there exist more degrees of freedom so that defects with (relatively) lower energy costs can be generated.

\item \textit{Activation energy:} We represent activation energy as a ratio of the low-temperature activation energy to the crossover temperature, which makes it a dimensionless number and independent of the system size. Whereas for the 2D system we have a ratio of $(0.26/(1/80)) \approx 21$, the corresponding ratio for the 3D system reads (data taken from reference \citenum{HeuerPhysRevLett2004}) (4.84 eV / 0.30 eV) $\approx$ 16. Thus, the effective barriers for the relaxation processes agree relatively well. The remaining differences may be related to a similar argument as in (iii) that in 3D there are more degrees of freedom which allow the system to explore more energy-efficient paths.

\end{enumerate}


\section{Structural aspects}
\label{sect:structure}

{In the low-temperature limit, where the system mainly resides close to the low-energy cutoff of defect-free states, the residual transitions between different defect-free states naturally involve the transient population defect states. This is supported by the observation that the low-temperature activation energy of 0.26 is significantly large than the difference in the cutoff energies of the states with and without defects. Furthermore, the deviations from the linear energy dependence vs. inverse temperature, expected for a Gaussian glass-former, start around the cutoff energy of defect states. For this purpose, we continue with the discussion of the structural properties of defect states, in particular close to their low-energy cutoff.}

\subsection{Ring statistics}

\begin{figure*}[!htb]
\centering
 \includegraphics[scale=1.0]{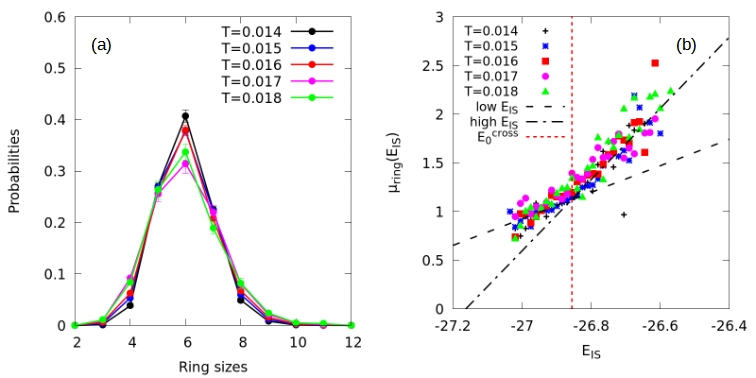}
 \caption{{(a) Probability distribution of the ring-sizes at various temperatures. (b) Dependence of the standard deviation of the ring distribution ($\mu_{ring}$) with energy for defect-free-states. Data below $E^{\text{cross}}_0$ and above $E^{\text{cross}}_0$ at T = 0.015 are fitted separately with a line.}}
 \label{fig:Energywise_varience}
\end{figure*}

{We start by studying the statistics of ring sizes as a function of energy. We found in Figure~\ref{fig:Energywise_varience}(a) that with increasing temperature, the probability of finding 5,6, and 7 rings decreases, whereas for all other rings it increases. It is an expected behavior, since large (> ringsize 7) and small (< ringsize 5) rings are quite strained and hence have higher energies~\cite{Roy_Heuer_PRL_2019,Roy_Heuer_JPhysCondMatt_2019}. In  Figure~\ref{fig:Energywise_varience}(b) we show the standard deviation of the ring distribution. As expected in thermodynamic equilibrium, at given energy the standard deviation does not depend on temperature within statistical uncertainties. One observes a different dependence on energy when crossing the crossover energy $E^{\text{cross}}_0$.}

\subsection{Energy-dependent defect properties}
\label{subsect:av_Defect}

\begin{figure*}[!htb]
 \includegraphics[scale=1]{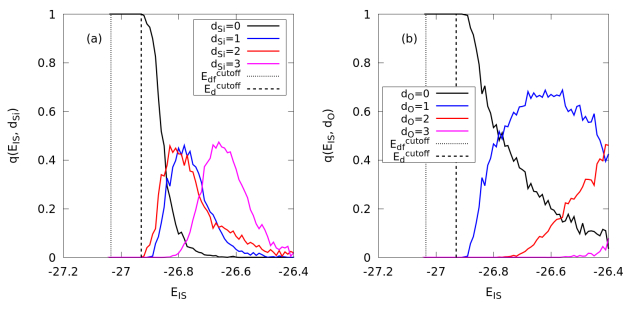}
 \caption{(a): Probability $q(E_{\text{IS}}, d_{\text{Si}})$ to find a specific defect state with $d_{\text{Si}}$ number of average defect-Si particles at energy $E_{\text{IS}}$.(b): Probability $q(E_{\text{IS}}, d_{\text{O}})$ to find a specific defect state with $d_{\text{O}}$ number of average defect-O particles at energy $E_{\text{IS}}$.}
 \label{fig:qav}
\end{figure*}

\begin{figure*}[!htb]
 \centering
 \includegraphics[scale=1.0]{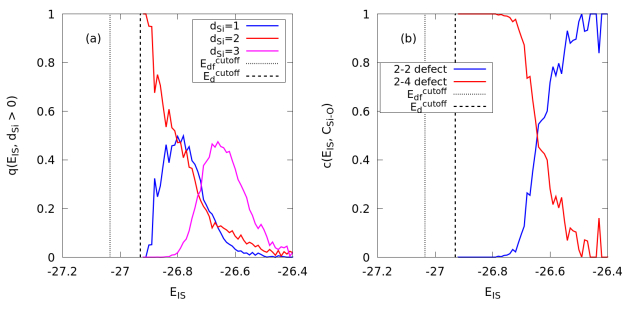}
 \caption{(a) Probability $q(E_{\text{IS}}, d_{\text{Si}}>0)$ to find a specific defect state with $d_{\text{Si}}$ number of average defect-Si particles at energy $E_{\text{IS}}$, excluding $d_{\text{Si}}=0$ states. (b) Probability $c(E_{\text{IS}}, C_{\text{Si-O}})$ to find a specific combination of the co-ordinations of the defect-Si particles for $d_{\text{Si}}=2$ states at energy $E_{\text{IS}}$, normalized by the total number of $d_{\text{Si}}=2$ states. 1-4 defects are ignored in this plot due to very small probability.}
 \label{fig:qav_2}
\end{figure*}

To examine the properties of the defect-states in the low-energy limit, we sort the defect-states according to the number of \textit{defect-Si} particles, denoted $d_{\text{Si}}$. Here, we do not distinguish whether a defect-Si particle has more or less than 3 O neighbors. The probabilities to find a specific number of defect-Si particles at a given energy are shown in Figure~\ref{fig:qav}(a). Clearly, the silicon defect states exactly disappear at $E^{\text{cutoff}}_{\text{d}}$. Again, the data are obtained from averaging over all temperatures, in analogy to Equation~\ref{eqn:average_fraction}. Actually, the average number of defect-Si particles linearly increases with energy above $E^{\text{cutoff}}_{\text{d}}$ (see supporting material, Figure SM.3). Similar to the observation of Yu et al~\cite{Yu_Wang_PhysRevMater_2021} in BKS-silica, we find that the slope of the number of defect particles vs. energy decreases with increasing energy. After excluding $d_{\text{Si}} = 0$ states, the appropriately renormalized data are shown in Figure~\ref{fig:qav_2}(a). We find that very close to $E^{\text{cutoff}}_{\text{d}}$, all defect-states have two silicon defects, i.e. $d_{\text{Si}} = 2$. In particular, there are no defect states with just one silicon defect. They only start to appear at slightly higher energies. This is also reflected in the anomalous behavior of the average number of defects (excluding $d_{\text{Si}} = 0$ states), shown in the supporting material Figure SM.3. 

Analogously, we study the properties of oxygen defects. In Figure~\ref{fig:qav}(b), the probabilities to find a specific number of defect-O particles are shown. We find that already above $E^{\text{cutoff}}_{\text{d}}$ no oxygen defects are present. Thus, there exists an energy range where two silicon defects are present but no oxygen defect (see below for a closer characterization of that state). For states slightly above $E^{\text{cutoff}}_{\text{d}}$, in average two defect-Si but no defect-O particles are present.

The structures of the $d_{\text{Si}} = 2$ defect states are further analyzed by counting different Si-O co-ordinations ($C_{\text{Si-O}}$) of the participating defect-Si particles $(i,j)$, such as, $C_{\text{Si-O}}^i=1$ and $C_{\text{Si-O}}^j=4$ (abbreviation: 1-4 defect); $(C_{\text{Si-O}}^i=2$ and $C_{\text{Si-O}}^j=2$ (abbreviation: 2-2 defect);$C_{\text{Si-O}}^i=2$ and $C_{\text{Si-O}}^j=4$ (abbreviation: 2-4 defect). As shown in Figure~\ref{fig:qav_2}(b), near $E^{\text{cutoff}}_{\text{d}}$ only 2-4 defect states are observed. Performing a similar analysis for $d_{\text{Si}}=1$ states, we find that only $C_{\text{Si-O}}=2$ co-ordination is present for the single defect-Si particle at all energies. We find that such a Si-O co-ordination for $d_{\text{Si}}=1$ states is not possible with $d_{\text{O}}=0$ but with $d_{\text{O}}=1$, as by definition co-ordination numbers of the Si and O particles are dependent on each other. Performing a similar co-ordination number analysis to only $d_{\text{O}}=1$ states, we find that corresponding $C_{\text{O-Si}}$ is strictly 1 within our simulation set up.

\begin{figure*}[!htb]
 \includegraphics[scale=1.0]{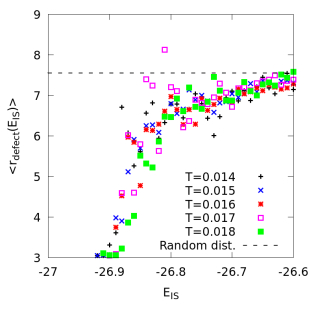}
 \caption{ Average distance between two defect-Si particles. The limit of a random distribution is shown at $<r_{\text{defect}}(E_{\text{IS}})> \sim 7.55$.}
 \label{fig:av_Defect_distance}
\end{figure*}

Next, we analyze the spatial correlations in between the defect-Si particles in each frame by calculating the average distance between them in dependence of energy, $r_{\text{defect}}(E_{\text{IS}})$. The results are shown in Figure~\ref{fig:av_Defect_distance}. We find that the minimum value is $<r_{\text{defect}}(E_{\text{IS}})> \sim 3$ {\AA}, which is also the average distance between two Si particles in a defect-free state~\cite{Roy_Heuer_PCCP_2018}. The average distance between the defect particles increases with energy and becomes constant after a certain energy range, suggesting that a breakdown of any spatial correlations among defect-Si particles. The limiting value of the average distance can be calculated by evaluating the integral,

\begin{equation}
\frac{L}{2} \int_0^1 dx \int_0^1 dy \sqrt{x^2 + y^2}  \sim 0.77\frac{L}{2}
\label{eqn:av_dist}
\end{equation}

which for the 80 particle system corresponds to $\sim 7.55$.

\subsection{Microscopic realizations}

\begin{figure*}[!htb]
\centering
 \includegraphics[scale=1.0]{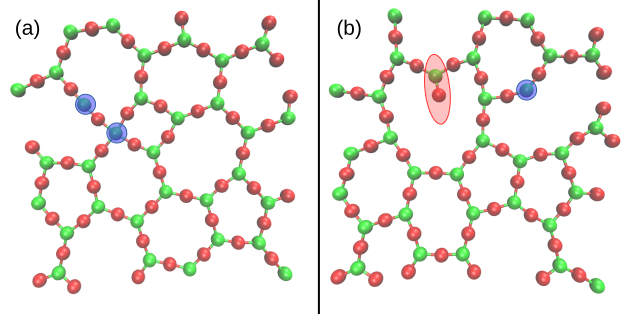}
 \caption{Sample structures for states with (a) 2 defect Si particles and (b) 1 defect Si particle. The defect Si particles are marked in blue. The dangling Si-O bond in (b) is marked with a red ellipse.}
 \label{fig:dSi_2_1}
 \end{figure*}

These results suggest that there exists a locally well-defined defect structure for the defect-states close to $E^{\text{cutoff}}_{\text{d}}$. This structure involves no defect-O particles but two close-by defect-Si particles within a distance of $\sim$ 3 {\AA}. The two defect-Si particles have a coordination number of 2 and 4, respectively, yielding the average co-ordination numbers $\langle C_{\text{Si-O}} \rangle=3$, $\langle C_{\text{O-Si}} \rangle=2$ for the total system at $E^{\text{cutoff}}_{\text{d}}$. 

A typical structure of the $d_{\text{Si}} = 2$ states close to the $E^{\text{cutoff}}_{\text{d}}$ is shown in Figure~\ref{fig:dSi_2_1}(a). Starting from a defect-free structure, one can generate this defect-structure by transferring one O particle from one Si particle to an adjacent Si particle. With increasing energy, more O particles are transferred further along the network and the average distance between defect-Si particles ($<r_{\text{defect}}>$) increases from 3 {\AA}. This result has a trivial size dependence, as the oxygen transfer can happen simultaneously at larger distances. As a result, the value of $<r_{\text{defect}}>$ in Figure~\ref{fig:av_Defect_distance} may increase at low energies for large systems. This again underlines the need to study small systems to clearly elucidate the energy-dependent structural properties at low energies. Furthermore, it explains our observation that  the difference between the two cutoff energies $\Delta E^{\text{cutoff}}$ is practically independent of system size. 

A typical state with $d_{\text{Si}} = 1$ is displayed in Figure~\ref{fig:dSi_2_1}(b). These structures involve a dangling O particle with coordination number 1. Comparison of the states with $d_{\text{Si}} = 2$ and $d_{\text{Si}} = 1$ suggest that from an energetic point of view a state with $d_{\text{Si}} = 2$ is closer to the defect-free states than $d_{\text{Si}} = 1$ and thus has the chance to display a lower energy. 

The result shows that the average number of defect particles and average co-ordination number of the system may not have a linear relationship with energy, at least near $E^{\text{cutoff}}_{\text{d}}$. Indeed, as shown in the supporting material Figure SM.4(a), the average co-ordination number of defect-Si states (excluding $d_{\text{Si}} = 0$ states) states shows a minimum, roughly around the same location where $d_{\text{Si}} = 1$ states are most populated. This effect can be attributed to the lower $\langle C_{\text{Si-O}} \rangle$ value of $d_{\text{Si}} = 1$ states as compared to the $d_{\text{Si}} = 2$ states. Interestingly, we find that for defect-O states (excluding $d_{\text{O}} = 0$ states), the average co-ordination number is constant to 1 throughout the relevant energy range within our simulation set up. The relatively late appearance of $d_{\text{O}} = 2,3$ states in the energy spectrum might be a reason for this observation. 


\section{Conclusion and Outlook}

In this work, we have shown that the phenomenon of FSC in 2D-silica model can be directly related to the properties of the PEL. We find that the both defect and defect-free-states have a low energy cutoff. The energy difference between these energies does not depend on system size which is a first hint that the low-energy defect-states are fully local. Similar to the case of BKS-silica, we find that at the temperature where the system senses the presence of the final low-energy cutoff, the fragile to strong crossover (FSC) in the kinetics is observed. The results suggest that the emergence of the low-energy cutoff of $G_{\text{df}}(E_{\text{IS}})$ (or $G(E_{\text{IS}})$) is the major cause of the FSC.

Structurally, the properties of the low-energy defect states have been clearly characterized. One can see that the defect ground-state is a spatially coupled pair of defect-Si particles which does not involve any dangling bond, in contrast to a structure with just a single defect-Si particle. Thus the absence of a dangling bond renders the double defect structure energetically more favorable.

{The qualitative similarity of the observed FSC phenomenon between 2D and 3D silica supports the notion that the FSC may be a generic phenomenon, observed for network-forming systems. Furthermore, it enables a quantitative comparison, revealing the impact of dimension. A major difference between both systems is seen in the normalized difference of the cutoff energies of the defect and defect-free states. As another difference, we have observed that the low-temperature activation energy for the 2D system is approx. 30\% higher than for the 3D system. Both effects may be related to the difficulty in the 2D system to find reaction coordinates with low energy barriers.} Furthermore, we would like to stress that the 2D activation energy is more than twice as high as the energy difference of the overall ground-state and the ground-state defect structure (0.26 vs. 0.11). This reflects the complexity which is involved in the transition between two defect-free states, dominating at low temperatures, as well as the additional impact of activation barriers.

{We would like to remark that here we observe a variety of different defect states (figure~\ref{fig:qav}) when increasing the energy somewhat above their cutoff energy. Furthermore, we found that for higher energies the density of states behaves like a Gaussian glass-forming system, as also supported by the linear dependence of $E_{\text{IS}}^{\text{av}}$ on $\beta$ in figure 2(b). A well-defined second Arrhenius-regime at higher temperatures is absent as well. Hence, a direct application of the two-state model~\cite{Tanaka_JPhysCondMatt_2003, Shi_Tanaka_PNAS_2018, Shi_Tanaka_JChemPhys_2018} does not seem to be possible.}

The present analysis has again shown that the choice of small system sizes is essential to learn about specific PEL-related properties. Otherwise the low-energy properties would be smeared out and it would be difficult to identify the microscopic origin of the FSC beyond the mere identification of similar crossover temperatures in the thermodynamic and kinetic properties.

It will be interesting to use a three-body force-field (like Stillinger-Weber potential \cite{StillingerPhysRevB1985}) in addition to the two-body Yukawa force-field, and analyze how the new 2D-silica model changes its properties near the FSC-regime. In the present 2D-silica model, it is possible to unambiguously separate defect and defect-free-states based on a single cutoff radius at all temperatures. It remains to be seen if similar analysis can be performed in other network forming materials which show FSC as well, such as water.  

\section*{SUPPLEMENTARY MATERIAL}
See supplementary material for (i) DOS curves for defect, defect-free, and total IS distributions; (ii) DOS curves for 80-particle and 200-particle systems; (iii) Average number of defect Si/O particles with energy; (iv) Average co-ordination number with energy. 

\section*{ACKNOWLEDGMENTS}
The authors thank the NRW Graduate School of Chemistry and the University of M\"unster for providing necessary funding. 

\section*{AUTHOR DECLARATIONS}

\subsection*{Conflict of Interest}
There are no conflicts of interests to declare. 

\subsection*{Author Contributions}
{\bf Andreas Heuer:} Conceptualization, funding acquisition, project administration, resource allocation, supervision, validation, and manuscript review. {\bf Projesh Kumar Roy:} data curation, formal analysis, software development, and visualization. Both authors contributed equally in writing the original draft of the manuscript. 

\section*{DATA AVAILABILITY}
The data that support the findings of this study are available from the corresponding author upon reasonable request.

\bibliographystyle{apsrev4-2}

\begin{thebibliography}{72}%
\makeatletter
\providecommand \@ifxundefined [1]{%
 \@ifx{#1\undefined}
}%
\providecommand \@ifnum [1]{%
 \ifnum #1\expandafter \@firstoftwo
 \else \expandafter \@secondoftwo
 \fi
}%
\providecommand \@ifx [1]{%
 \ifx #1\expandafter \@firstoftwo
 \else \expandafter \@secondoftwo
 \fi
}%
\providecommand \natexlab [1]{#1}%
\providecommand \enquote  [1]{``#1''}%
\providecommand \bibnamefont  [1]{#1}%
\providecommand \bibfnamefont [1]{#1}%
\providecommand \citenamefont [1]{#1}%
\providecommand \href@noop [0]{\@secondoftwo}%
\providecommand \href [0]{\begingroup \@sanitize@url \@href}%
\providecommand \@href[1]{\@@startlink{#1}\@@href}%
\providecommand \@@href[1]{\endgroup#1\@@endlink}%
\providecommand \@sanitize@url [0]{\catcode `\\12\catcode `\$12\catcode
  `\&12\catcode `\#12\catcode `\^12\catcode `\_12\catcode `\%12\relax}%
\providecommand \@@startlink[1]{}%
\providecommand \@@endlink[0]{}%
\providecommand \url  [0]{\begingroup\@sanitize@url \@url }%
\providecommand \@url [1]{\endgroup\@href {#1}{\urlprefix }}%
\providecommand \urlprefix  [0]{URL }%
\providecommand \Eprint [0]{\href }%
\providecommand \doibase [0]{https://doi.org/}%
\providecommand \selectlanguage [0]{\@gobble}%
\providecommand \bibinfo  [0]{\@secondoftwo}%
\providecommand \bibfield  [0]{\@secondoftwo}%
\providecommand \translation [1]{[#1]}%
\providecommand \BibitemOpen [0]{}%
\providecommand \bibitemStop [0]{}%
\providecommand \bibitemNoStop [0]{.\EOS\space}%
\providecommand \EOS [0]{\spacefactor3000\relax}%
\providecommand \BibitemShut  [1]{\csname bibitem#1\endcsname}%
\let\auto@bib@innerbib\@empty
\bibitem [{\citenamefont {Angell}(1991)}]{AngellJNonCrysSolids1991}%
  \BibitemOpen
  \bibfield  {author} {\bibinfo {author} {\bibfnamefont {C.~A.}\ \bibnamefont
  {Angell}},\ }\href
  {https://doi.org/https://doi.org/10.1016/0022-3093(91)90266-9} {\bibfield
  {journal} {\bibinfo  {journal} {Journal of Non-Crystalline Solids}\ }\textbf
  {\bibinfo {volume} {131-133}},\ \bibinfo {pages} {13 } (\bibinfo {year}
  {1991})}\BibitemShut {NoStop}%
\bibitem [{\citenamefont {Einstein}(1906)}]{Einstein_AnnPhysik_1906}%
  \BibitemOpen
  \bibfield  {author} {\bibinfo {author} {\bibfnamefont {A.}~\bibnamefont
  {Einstein}},\ }\href@noop {} {\bibfield  {journal} {\bibinfo  {journal} {Ann.
  Physik}\ }\textbf {\bibinfo {volume} {19}},\ \bibinfo {pages} {371} (\bibinfo
  {year} {1906})}\BibitemShut {NoStop}%
\bibitem [{\citenamefont {Vogel}(1921)}]{VogelPhysZ1921}%
  \BibitemOpen
  \bibfield  {author} {\bibinfo {author} {\bibfnamefont {H.}~\bibnamefont
  {Vogel}},\ }\href@noop {} {\bibfield  {journal} {\bibinfo  {journal} {Phys.
  Z.}\ }\textbf {\bibinfo {volume} {22}},\ \bibinfo {pages} {645} (\bibinfo
  {year} {1921})}\BibitemShut {NoStop}%
\bibitem [{\citenamefont {Fulcher}(1925)}]{FuchlerJAmCeramSoc1925}%
  \BibitemOpen
  \bibfield  {author} {\bibinfo {author} {\bibfnamefont {G.~S.}\ \bibnamefont
  {Fulcher}},\ }\href {https://doi.org/10.1111/j.1151-2916.1925.tb16731.x}
  {\bibfield  {journal} {\bibinfo  {journal} {Journal of the American Ceramic
  Society}\ }\textbf {\bibinfo {volume} {8}},\ \bibinfo {pages} {339} (\bibinfo
  {year} {1925})}\BibitemShut {NoStop}%
\bibitem [{\citenamefont {Tamman}\ and\ \citenamefont
  {Hesse}(1926)}]{TammanZAnorgAllgChem1926}%
  \BibitemOpen
  \bibfield  {author} {\bibinfo {author} {\bibfnamefont {G.}~\bibnamefont
  {Tamman}}\ and\ \bibinfo {author} {\bibfnamefont {W.}~\bibnamefont {Hesse}},\
  }\href@noop {} {\bibfield  {journal} {\bibinfo  {journal} {Z. Anorg. Allg.
  Chem.}\ }\textbf {\bibinfo {volume} {156}},\ \bibinfo {pages} {245} (\bibinfo
  {year} {1926})}\BibitemShut {NoStop}%
\bibitem [{\citenamefont {Scherer}(1992)}]{VFTJAmCerSoc1992}%
  \BibitemOpen
  \bibfield  {author} {\bibinfo {author} {\bibfnamefont {G.~W.}\ \bibnamefont
  {Scherer}},\ }\href {https://doi.org/10.1111/j.1151-2916.1992.tb05537.x}
  {\bibfield  {journal} {\bibinfo  {journal} {Journal of the American Ceramic
  Society}\ }\textbf {\bibinfo {volume} {75}},\ \bibinfo {pages} {1060}
  (\bibinfo {year} {1992})}\BibitemShut {NoStop}%
\bibitem [{\citenamefont {Angell}(1983)}]{Angell_AnnRevPhysChem_1983}%
  \BibitemOpen
  \bibfield  {author} {\bibinfo {author} {\bibfnamefont {C.~A.}\ \bibnamefont
  {Angell}},\ }\href {https://doi.org/10.1146/annurev.pc.34.100183.003113}
  {\bibfield  {journal} {\bibinfo  {journal} {Annual Review of Physical
  Chemistry}\ }\textbf {\bibinfo {volume} {34}},\ \bibinfo {pages} {593}
  (\bibinfo {year} {1983})}\BibitemShut {NoStop}%
\bibitem [{\citenamefont {Angell}(1995)}]{Angell_Science_1995}%
  \BibitemOpen
  \bibfield  {author} {\bibinfo {author} {\bibfnamefont {C.~A.}\ \bibnamefont
  {Angell}},\ }\href {https://doi.org/10.1126/science.267.5206.1924} {\bibfield
   {journal} {\bibinfo  {journal} {Science}\ }\textbf {\bibinfo {volume}
  {267}},\ \bibinfo {pages} {1924} (\bibinfo {year} {1995})}\BibitemShut
  {NoStop}%
\bibitem [{\citenamefont {Hess}\ \emph {et~al.}(1996)\citenamefont {Hess},
  \citenamefont {Dingwell},\ and\ \citenamefont
  {RÃ¶ssler}}]{Hess_Rossler_ChemGeo_1996}%
  \BibitemOpen
  \bibfield  {author} {\bibinfo {author} {\bibfnamefont {K.-U.}\ \bibnamefont
  {Hess}}, \bibinfo {author} {\bibfnamefont {D.}~\bibnamefont {Dingwell}},\
  and\ \bibinfo {author} {\bibfnamefont {E.}~\bibnamefont {RÃ¶ssler}},\ }\href
  {https://doi.org/https://doi.org/10.1016/0009-2541(95)00170-0} {\bibfield
  {journal} {\bibinfo  {journal} {Chemical Geology}\ }\textbf {\bibinfo
  {volume} {128}},\ \bibinfo {pages} {155} (\bibinfo {year} {1996})},\ \bibinfo
  {note} {5TH Silicate Melt Workshop}\BibitemShut {NoStop}%
\bibitem [{\citenamefont {Hemmati}\ \emph {et~al.}(2001)\citenamefont
  {Hemmati}, \citenamefont {Moynihan},\ and\ \citenamefont
  {Angell}}]{Hemmati_Angell_JChemPhys_2001}%
  \BibitemOpen
  \bibfield  {author} {\bibinfo {author} {\bibfnamefont {M.}~\bibnamefont
  {Hemmati}}, \bibinfo {author} {\bibfnamefont {C.~T.}\ \bibnamefont
  {Moynihan}},\ and\ \bibinfo {author} {\bibfnamefont {C.~A.}\ \bibnamefont
  {Angell}},\ }\href {https://doi.org/10.1063/1.1396679} {\bibfield  {journal}
  {\bibinfo  {journal} {The Journal of Chemical Physics}\ }\textbf {\bibinfo
  {volume} {115}},\ \bibinfo {pages} {6663} (\bibinfo {year}
  {2001})}\BibitemShut {NoStop}%
\bibitem [{\citenamefont {Riebling}(1965)}]{Reibling_JChemPhys_1965}%
  \BibitemOpen
  \bibfield  {author} {\bibinfo {author} {\bibfnamefont {E.~F.}\ \bibnamefont
  {Riebling}},\ }\href {https://doi.org/10.1063/1.1696771} {\bibfield
  {journal} {\bibinfo  {journal} {The Journal of Chemical Physics}\ }\textbf
  {\bibinfo {volume} {43}},\ \bibinfo {pages} {499} (\bibinfo {year}
  {1965})}\BibitemShut {NoStop}%
\bibitem [{\citenamefont {Lucas}\ \emph {et~al.}(2017)\citenamefont {Lucas},
  \citenamefont {Coleman}, \citenamefont {Venkateswara~Rao}, \citenamefont
  {Edwards}, \citenamefont {Devaadithya}, \citenamefont {Wei}, \citenamefont
  {Alsayoud}, \citenamefont {Potter}, \citenamefont {Muralidharan},\ and\
  \citenamefont {Deymier}}]{Lucas_Deymier_JPhysChemB_2017}%
  \BibitemOpen
  \bibfield  {author} {\bibinfo {author} {\bibfnamefont {P.}~\bibnamefont
  {Lucas}}, \bibinfo {author} {\bibfnamefont {G.~J.}\ \bibnamefont {Coleman}},
  \bibinfo {author} {\bibfnamefont {M.}~\bibnamefont {Venkateswara~Rao}},
  \bibinfo {author} {\bibfnamefont {A.~N.}\ \bibnamefont {Edwards}}, \bibinfo
  {author} {\bibfnamefont {C.}~\bibnamefont {Devaadithya}}, \bibinfo {author}
  {\bibfnamefont {S.}~\bibnamefont {Wei}}, \bibinfo {author} {\bibfnamefont
  {A.~Q.}\ \bibnamefont {Alsayoud}}, \bibinfo {author} {\bibfnamefont {B.~G.}\
  \bibnamefont {Potter}}, \bibinfo {author} {\bibfnamefont {K.}~\bibnamefont
  {Muralidharan}},\ and\ \bibinfo {author} {\bibfnamefont {P.~A.}\ \bibnamefont
  {Deymier}},\ }\href {https://doi.org/10.1021/acs.jpcb.7b10857} {\bibfield
  {journal} {\bibinfo  {journal} {The Journal of Physical Chemistry B}\
  }\textbf {\bibinfo {volume} {121}},\ \bibinfo {pages} {11210} (\bibinfo
  {year} {2017})},\ \bibinfo {note} {pMID: 29166015}\BibitemShut {NoStop}%
\bibitem [{\citenamefont {Lucas}(2019)}]{Lucas_JNonCrysSolidsX_2019}%
  \BibitemOpen
  \bibfield  {author} {\bibinfo {author} {\bibfnamefont {P.}~\bibnamefont
  {Lucas}},\ }\href
  {https://doi.org/https://doi.org/10.1016/j.nocx.2019.100034} {\bibfield
  {journal} {\bibinfo  {journal} {Journal of Non-Crystalline Solids: X}\
  }\textbf {\bibinfo {volume} {4}},\ \bibinfo {pages} {100034} (\bibinfo {year}
  {2019})}\BibitemShut {NoStop}%
\bibitem [{\citenamefont {Shi}\ and\ \citenamefont
  {Tanaka}(2018{\natexlab{a}})}]{Shi_Tanaka_PNAS_2017}%
  \BibitemOpen
  \bibfield  {author} {\bibinfo {author} {\bibfnamefont {R.}~\bibnamefont
  {Shi}}\ and\ \bibinfo {author} {\bibfnamefont {H.}~\bibnamefont {Tanaka}},\
  }\href {https://doi.org/10.1073/pnas.1717233115} {\bibfield  {journal}
  {\bibinfo  {journal} {Proceedings of the National Academy of Sciences}\
  }\textbf {\bibinfo {volume} {115}},\ \bibinfo {pages} {1980} (\bibinfo {year}
  {2018}{\natexlab{a}})}\BibitemShut {NoStop}%
\bibitem [{\citenamefont {Russo}\ \emph {et~al.}(2018)\citenamefont {Russo},
  \citenamefont {Akahane},\ and\ \citenamefont
  {Tanaka}}]{Russo_Tanaka_PNAS_2018}%
  \BibitemOpen
  \bibfield  {author} {\bibinfo {author} {\bibfnamefont {J.}~\bibnamefont
  {Russo}}, \bibinfo {author} {\bibfnamefont {K.}~\bibnamefont {Akahane}},\
  and\ \bibinfo {author} {\bibfnamefont {H.}~\bibnamefont {Tanaka}},\ }\href
  {https://doi.org/10.1073/pnas.1722339115} {\bibfield  {journal} {\bibinfo
  {journal} {Proceedings of the National Academy of Sciences}\ }\textbf
  {\bibinfo {volume} {115}},\ \bibinfo {pages} {E3333} (\bibinfo {year}
  {2018})}\BibitemShut {NoStop}%
\bibitem [{\citenamefont {Shi}\ and\ \citenamefont
  {Tanaka}(2019)}]{Shi_Tanaka_SciAdv_2019}%
  \BibitemOpen
  \bibfield  {author} {\bibinfo {author} {\bibfnamefont {R.}~\bibnamefont
  {Shi}}\ and\ \bibinfo {author} {\bibfnamefont {H.}~\bibnamefont {Tanaka}},\
  }\bibfield  {journal} {\bibinfo  {journal} {Science Advances}\ }\textbf
  {\bibinfo {volume} {5}},\ \href {https://doi.org/10.1126/sciadv.aav3194}
  {10.1126/sciadv.aav3194} (\bibinfo {year} {2019})\BibitemShut {NoStop}%
\bibitem [{\citenamefont {De~Marzio}\ \emph {et~al.}(2016)\citenamefont
  {De~Marzio}, \citenamefont {Camisasca}, \citenamefont {Rovere},\ and\
  \citenamefont {Gallo}}]{Marzio_Gallo_JChemPhys_2016}%
  \BibitemOpen
  \bibfield  {author} {\bibinfo {author} {\bibfnamefont {M.}~\bibnamefont
  {De~Marzio}}, \bibinfo {author} {\bibfnamefont {G.}~\bibnamefont
  {Camisasca}}, \bibinfo {author} {\bibfnamefont {M.}~\bibnamefont {Rovere}},\
  and\ \bibinfo {author} {\bibfnamefont {P.}~\bibnamefont {Gallo}},\ }\href
  {https://doi.org/10.1063/1.4941946} {\bibfield  {journal} {\bibinfo
  {journal} {The Journal of Chemical Physics}\ }\textbf {\bibinfo {volume}
  {144}},\ \bibinfo {pages} {074503} (\bibinfo {year} {2016})}\BibitemShut
  {NoStop}%
\bibitem [{\citenamefont {Starr}\ \emph {et~al.}(1999)\citenamefont {Starr},
  \citenamefont {Sciortino},\ and\ \citenamefont
  {Stanley}}]{Starr_Stanley_PhysRevE_1999}%
  \BibitemOpen
  \bibfield  {author} {\bibinfo {author} {\bibfnamefont {F.~W.}\ \bibnamefont
  {Starr}}, \bibinfo {author} {\bibfnamefont {F.}~\bibnamefont {Sciortino}},\
  and\ \bibinfo {author} {\bibfnamefont {H.~E.}\ \bibnamefont {Stanley}},\
  }\href {https://doi.org/10.1103/PhysRevE.60.6757} {\bibfield  {journal}
  {\bibinfo  {journal} {Phys. Rev. E}\ }\textbf {\bibinfo {volume} {60}},\
  \bibinfo {pages} {6757} (\bibinfo {year} {1999})}\BibitemShut {NoStop}%
\bibitem [{\citenamefont {Xu}\ \emph {et~al.}(2005)\citenamefont {Xu},
  \citenamefont {Kumar}, \citenamefont {Buldyrev}, \citenamefont {Chen},
  \citenamefont {Poole}, \citenamefont {Sciortino},\ and\ \citenamefont
  {Stanley}}]{Xu_PNAS_2005}%
  \BibitemOpen
  \bibfield  {author} {\bibinfo {author} {\bibfnamefont {L.}~\bibnamefont
  {Xu}}, \bibinfo {author} {\bibfnamefont {P.}~\bibnamefont {Kumar}}, \bibinfo
  {author} {\bibfnamefont {S.~V.}\ \bibnamefont {Buldyrev}}, \bibinfo {author}
  {\bibfnamefont {S.-H.}\ \bibnamefont {Chen}}, \bibinfo {author}
  {\bibfnamefont {P.~H.}\ \bibnamefont {Poole}}, \bibinfo {author}
  {\bibfnamefont {F.}~\bibnamefont {Sciortino}},\ and\ \bibinfo {author}
  {\bibfnamefont {H.~E.}\ \bibnamefont {Stanley}},\ }\href
  {https://doi.org/10.1073/pnas.0507870102} {\bibfield  {journal} {\bibinfo
  {journal} {Proceedings of the National Academy of Sciences}\ }\textbf
  {\bibinfo {volume} {102}},\ \bibinfo {pages} {16558} (\bibinfo {year}
  {2005})}\BibitemShut {NoStop}%
\bibitem [{\citenamefont {Poole}\ \emph {et~al.}(1992)\citenamefont {Poole},
  \citenamefont {Sciortino}, \citenamefont {Essmann},\ and\ \citenamefont
  {Stanley}}]{Poole_Stanley_Nature_1992}%
  \BibitemOpen
  \bibfield  {author} {\bibinfo {author} {\bibfnamefont {P.~H.}\ \bibnamefont
  {Poole}}, \bibinfo {author} {\bibfnamefont {F.}~\bibnamefont {Sciortino}},
  \bibinfo {author} {\bibfnamefont {U.}~\bibnamefont {Essmann}},\ and\ \bibinfo
  {author} {\bibfnamefont {H.~E.}\ \bibnamefont {Stanley}},\ }\href
  {https://doi.org/10.1038/360324a0} {\bibfield  {journal} {\bibinfo  {journal}
  {Nature}\ }\textbf {\bibinfo {volume} {360}},\ \bibinfo {pages} {324}
  (\bibinfo {year} {1992})}\BibitemShut {NoStop}%
\bibitem [{\citenamefont {Mishima}\ and\ \citenamefont
  {Stanley}(1998)}]{Mishima_Stanley_Nature_1998}%
  \BibitemOpen
  \bibfield  {author} {\bibinfo {author} {\bibfnamefont {O.}~\bibnamefont
  {Mishima}}\ and\ \bibinfo {author} {\bibfnamefont {H.~E.}\ \bibnamefont
  {Stanley}},\ }\href {https://doi.org/10.1038/24540} {\bibfield  {journal}
  {\bibinfo  {journal} {Nature}\ }\textbf {\bibinfo {volume} {396}},\ \bibinfo
  {pages} {329} (\bibinfo {year} {1998})}\BibitemShut {NoStop}%
\bibitem [{\citenamefont {Palmer}\ \emph {et~al.}(2014)\citenamefont {Palmer},
  \citenamefont {Martelli}, \citenamefont {Liu}, \citenamefont {Car},
  \citenamefont {Panagiotopoulos},\ and\ \citenamefont
  {Debenedetti}}]{Palmer_Debenedetti_Nature_2014}%
  \BibitemOpen
  \bibfield  {author} {\bibinfo {author} {\bibfnamefont {J.~C.}\ \bibnamefont
  {Palmer}}, \bibinfo {author} {\bibfnamefont {F.}~\bibnamefont {Martelli}},
  \bibinfo {author} {\bibfnamefont {Y.}~\bibnamefont {Liu}}, \bibinfo {author}
  {\bibfnamefont {R.}~\bibnamefont {Car}}, \bibinfo {author} {\bibfnamefont
  {A.~Z.}\ \bibnamefont {Panagiotopoulos}},\ and\ \bibinfo {author}
  {\bibfnamefont {P.~G.}\ \bibnamefont {Debenedetti}},\ }\href
  {https://doi.org/10.1038/nature13405} {\bibfield  {journal} {\bibinfo
  {journal} {Nature}\ }\textbf {\bibinfo {volume} {510}},\ \bibinfo {pages}
  {385} (\bibinfo {year} {2014})}\BibitemShut {NoStop}%
\bibitem [{\citenamefont {Poole}\ \emph
  {et~al.}(1993{\natexlab{a}})\citenamefont {Poole}, \citenamefont {Sciortino},
  \citenamefont {Essmann},\ and\ \citenamefont
  {Stanley}}]{Poole_Stanley_PhysRevE_1993}%
  \BibitemOpen
  \bibfield  {author} {\bibinfo {author} {\bibfnamefont {P.~H.}\ \bibnamefont
  {Poole}}, \bibinfo {author} {\bibfnamefont {F.}~\bibnamefont {Sciortino}},
  \bibinfo {author} {\bibfnamefont {U.}~\bibnamefont {Essmann}},\ and\ \bibinfo
  {author} {\bibfnamefont {H.~E.}\ \bibnamefont {Stanley}},\ }\href
  {https://doi.org/10.1103/PhysRevE.48.3799} {\bibfield  {journal} {\bibinfo
  {journal} {Phys. Rev. E}\ }\textbf {\bibinfo {volume} {48}},\ \bibinfo
  {pages} {3799} (\bibinfo {year} {1993}{\natexlab{a}})}\BibitemShut {NoStop}%
\bibitem [{\citenamefont {Poole}\ \emph
  {et~al.}(1993{\natexlab{b}})\citenamefont {Poole}, \citenamefont {Essmann},
  \citenamefont {Sciortino},\ and\ \citenamefont
  {Stanley}}]{Poole_Stanley_PhysRevE_1993_2}%
  \BibitemOpen
  \bibfield  {author} {\bibinfo {author} {\bibfnamefont {P.~H.}\ \bibnamefont
  {Poole}}, \bibinfo {author} {\bibfnamefont {U.}~\bibnamefont {Essmann}},
  \bibinfo {author} {\bibfnamefont {F.}~\bibnamefont {Sciortino}},\ and\
  \bibinfo {author} {\bibfnamefont {H.~E.}\ \bibnamefont {Stanley}},\ }\href
  {https://doi.org/10.1103/PhysRevE.48.4605} {\bibfield  {journal} {\bibinfo
  {journal} {Phys. Rev. E}\ }\textbf {\bibinfo {volume} {48}},\ \bibinfo
  {pages} {4605} (\bibinfo {year} {1993}{\natexlab{b}})}\BibitemShut {NoStop}%
\bibitem [{\citenamefont {Stillinger}\ and\ \citenamefont
  {Weber}(1982)}]{Stillinger_Weber_PhysRevA_1982}%
  \BibitemOpen
  \bibfield  {author} {\bibinfo {author} {\bibfnamefont {F.~H.}\ \bibnamefont
  {Stillinger}}\ and\ \bibinfo {author} {\bibfnamefont {T.~A.}\ \bibnamefont
  {Weber}},\ }\href {https://doi.org/10.1103/PhysRevA.25.978} {\bibfield
  {journal} {\bibinfo  {journal} {Phys Rev A}\ }\textbf {\bibinfo {volume}
  {25}},\ \bibinfo {pages} {978} (\bibinfo {year} {1982})}\BibitemShut
  {NoStop}%
\bibitem [{\citenamefont {Stillinger}\ and\ \citenamefont
  {Weber}(1983{\natexlab{a}})}]{Stillinger_Weber_PhysRevA_1983}%
  \BibitemOpen
  \bibfield  {author} {\bibinfo {author} {\bibfnamefont {F.~H.}\ \bibnamefont
  {Stillinger}}\ and\ \bibinfo {author} {\bibfnamefont {T.~A.}\ \bibnamefont
  {Weber}},\ }\href {https://doi.org/10.1103/PhysRevA.28.2408} {\bibfield
  {journal} {\bibinfo  {journal} {Phys. Rev. A}\ }\textbf {\bibinfo {volume}
  {28}},\ \bibinfo {pages} {2408} (\bibinfo {year}
  {1983}{\natexlab{a}})}\BibitemShut {NoStop}%
\bibitem [{\citenamefont {Stillinger}\ and\ \citenamefont
  {Weber}(1984)}]{Stillinger_Weber_Science_1984}%
  \BibitemOpen
  \bibfield  {author} {\bibinfo {author} {\bibfnamefont {F.~H.}\ \bibnamefont
  {Stillinger}}\ and\ \bibinfo {author} {\bibfnamefont {T.~A.}\ \bibnamefont
  {Weber}},\ }\href {https://doi.org/10.1126/science.225.4666.983} {\bibfield
  {journal} {\bibinfo  {journal} {Science}\ }\textbf {\bibinfo {volume}
  {225}},\ \bibinfo {pages} {983} (\bibinfo {year} {1984})}\BibitemShut
  {NoStop}%
\bibitem [{\citenamefont {Stillinger}\ and\ \citenamefont
  {Weber}(1983{\natexlab{b}})}]{Stillinger_Weber_JPhysChem_1983}%
  \BibitemOpen
  \bibfield  {author} {\bibinfo {author} {\bibfnamefont {F.~H.}\ \bibnamefont
  {Stillinger}}\ and\ \bibinfo {author} {\bibfnamefont {T.~A.}\ \bibnamefont
  {Weber}},\ }\href {https://doi.org/10.1021/j100238a027} {\bibfield  {journal}
  {\bibinfo  {journal} {J. Phys. Chem.}\ }\textbf {\bibinfo {volume} {87}},\
  \bibinfo {pages} {2833} (\bibinfo {year} {1983}{\natexlab{b}})}\BibitemShut
  {NoStop}%
\bibitem [{\citenamefont {Weber}\ and\ \citenamefont
  {Stillinger}(1984)}]{Weber_Stillinger_JChemPhys_1984}%
  \BibitemOpen
  \bibfield  {author} {\bibinfo {author} {\bibfnamefont {T.~A.}\ \bibnamefont
  {Weber}}\ and\ \bibinfo {author} {\bibfnamefont {F.~H.}\ \bibnamefont
  {Stillinger}},\ }\href {https://doi.org/10.1063/1.447072} {\bibfield
  {journal} {\bibinfo  {journal} {J. Chem. Phys.}\ }\textbf {\bibinfo {volume}
  {80}},\ \bibinfo {pages} {2742} (\bibinfo {year} {1984})}\BibitemShut
  {NoStop}%
\bibitem [{\citenamefont {Sastry}\ \emph {et~al.}(1998)\citenamefont {Sastry},
  \citenamefont {Debenedetti},\ and\ \citenamefont
  {Stillinger}}]{Sastry_Stillinger_Nature_1998}%
  \BibitemOpen
  \bibfield  {author} {\bibinfo {author} {\bibfnamefont {S.}~\bibnamefont
  {Sastry}}, \bibinfo {author} {\bibfnamefont {P.~G.}\ \bibnamefont
  {Debenedetti}},\ and\ \bibinfo {author} {\bibfnamefont {F.~H.}\ \bibnamefont
  {Stillinger}},\ }\href {https://doi.org/10.1038/31189} {\bibfield  {journal}
  {\bibinfo  {journal} {Nature}\ }\textbf {\bibinfo {volume} {393}},\ \bibinfo
  {pages} {554} (\bibinfo {year} {1998})}\BibitemShut {NoStop}%
\bibitem [{\citenamefont {Debenedetti}\ and\ \citenamefont
  {Stillinger}(2001)}]{Debnedetti_Stillinger_Nature_2001}%
  \BibitemOpen
  \bibfield  {author} {\bibinfo {author} {\bibfnamefont {P.~G.}\ \bibnamefont
  {Debenedetti}}\ and\ \bibinfo {author} {\bibfnamefont {F.~H.}\ \bibnamefont
  {Stillinger}},\ }\href {https://doi.org/10.1038/35065704} {\bibfield
  {journal} {\bibinfo  {journal} {Nature}\ }\textbf {\bibinfo {volume} {410}},\
  \bibinfo {pages} {259} (\bibinfo {year} {2001})}\BibitemShut {NoStop}%
\bibitem [{\citenamefont {Heuer}(2008)}]{Heuer_JPhysCondMatt_2008}%
  \BibitemOpen
  \bibfield  {author} {\bibinfo {author} {\bibfnamefont {A.}~\bibnamefont
  {Heuer}},\ }\href@noop {} {\bibfield  {journal} {\bibinfo  {journal} {Journal
  of Physics: Condensed Matter}\ }\textbf {\bibinfo {volume} {20}},\ \bibinfo
  {pages} {373101} (\bibinfo {year} {2008})}\BibitemShut {NoStop}%
\bibitem [{\citenamefont {Sastry}(2001)}]{Sastry_Nature_2001}%
  \BibitemOpen
  \bibfield  {author} {\bibinfo {author} {\bibfnamefont {S.}~\bibnamefont
  {Sastry}},\ }\href {https://doi.org/10.1038/35051524} {\bibfield  {journal}
  {\bibinfo  {journal} {Nature}\ }\textbf {\bibinfo {volume} {409}},\ \bibinfo
  {pages} {164} (\bibinfo {year} {2001})}\BibitemShut {NoStop}%
\bibitem [{\citenamefont {Martinez}\ and\ \citenamefont
  {Angell}(2001)}]{Martinez_Angell_Nature_2001}%
  \BibitemOpen
  \bibfield  {author} {\bibinfo {author} {\bibfnamefont {L.-M.}\ \bibnamefont
  {Martinez}}\ and\ \bibinfo {author} {\bibfnamefont {C.~A.}\ \bibnamefont
  {Angell}},\ }\href {https://doi.org/10.1038/35070517} {\bibfield  {journal}
  {\bibinfo  {journal} {Nature}\ }\textbf {\bibinfo {volume} {410}},\ \bibinfo
  {pages} {663} (\bibinfo {year} {2001})}\BibitemShut {NoStop}%
\bibitem [{\citenamefont {Adam}\ and\ \citenamefont
  {Gibbs}(1965)}]{Adam_Gibbs_JChemPhys_1965}%
  \BibitemOpen
  \bibfield  {author} {\bibinfo {author} {\bibfnamefont {G.}~\bibnamefont
  {Adam}}\ and\ \bibinfo {author} {\bibfnamefont {J.~H.}\ \bibnamefont
  {Gibbs}},\ }\href {https://doi.org/10.1063/1.1696442} {\bibfield  {journal}
  {\bibinfo  {journal} {The Journal of Chemical Physics}\ }\textbf {\bibinfo
  {volume} {43}},\ \bibinfo {pages} {139} (\bibinfo {year} {1965})}\BibitemShut
  {NoStop}%
\bibitem [{\citenamefont {Ito}\ \emph {et~al.}(1999)\citenamefont {Ito},
  \citenamefont {Moynihan},\ and\ \citenamefont
  {Angell}}]{Kaori_Angell_Nature_1999}%
  \BibitemOpen
  \bibfield  {author} {\bibinfo {author} {\bibfnamefont {K.}~\bibnamefont
  {Ito}}, \bibinfo {author} {\bibfnamefont {C.~T.}\ \bibnamefont {Moynihan}},\
  and\ \bibinfo {author} {\bibfnamefont {C.~A.}\ \bibnamefont {Angell}},\
  }\href {https://doi.org/10.1038/19042} {\bibfield  {journal} {\bibinfo
  {journal} {Nature}\ }\textbf {\bibinfo {volume} {398}},\ \bibinfo {pages}
  {492} (\bibinfo {year} {1999})}\BibitemShut {NoStop}%
\bibitem [{\citenamefont {Saika-Voivod}\ \emph {et~al.}(2001)\citenamefont
  {Saika-Voivod}, \citenamefont {Poole},\ and\ \citenamefont
  {Sciortino}}]{Voivod_Sciortino_Nature_2001}%
  \BibitemOpen
  \bibfield  {author} {\bibinfo {author} {\bibfnamefont {I.}~\bibnamefont
  {Saika-Voivod}}, \bibinfo {author} {\bibfnamefont {P.~H.}\ \bibnamefont
  {Poole}},\ and\ \bibinfo {author} {\bibfnamefont {F.}~\bibnamefont
  {Sciortino}},\ }\href {https://doi.org/10.1038/35087524} {\bibfield
  {journal} {\bibinfo  {journal} {Nature}\ }\textbf {\bibinfo {volume} {412}},\
  \bibinfo {pages} {514} (\bibinfo {year} {2001})}\BibitemShut {NoStop}%
\bibitem [{\citenamefont {Saika-Voivod}\ \emph {et~al.}(2004)\citenamefont
  {Saika-Voivod}, \citenamefont {Sciortino},\ and\ \citenamefont
  {Poole}}]{Voivod_Poole_PhysRevE_2004}%
  \BibitemOpen
  \bibfield  {author} {\bibinfo {author} {\bibfnamefont {I.}~\bibnamefont
  {Saika-Voivod}}, \bibinfo {author} {\bibfnamefont {F.}~\bibnamefont
  {Sciortino}},\ and\ \bibinfo {author} {\bibfnamefont {P.~H.}\ \bibnamefont
  {Poole}},\ }\href {https://doi.org/10.1103/PhysRevE.69.041503} {\bibfield
  {journal} {\bibinfo  {journal} {Phys. Rev. E}\ }\textbf {\bibinfo {volume}
  {69}},\ \bibinfo {pages} {041503} (\bibinfo {year} {2004})}\BibitemShut
  {NoStop}%
\bibitem [{\citenamefont {Saksaengwijit}\ \emph {et~al.}(2004)\citenamefont
  {Saksaengwijit}, \citenamefont {Reinisch},\ and\ \citenamefont
  {Heuer}}]{HeuerPhysRevLett2004}%
  \BibitemOpen
  \bibfield  {author} {\bibinfo {author} {\bibfnamefont {A.}~\bibnamefont
  {Saksaengwijit}}, \bibinfo {author} {\bibfnamefont {J.}~\bibnamefont
  {Reinisch}},\ and\ \bibinfo {author} {\bibfnamefont {A.}~\bibnamefont
  {Heuer}},\ }\href {https://doi.org/10.1103/PhysRevLett.93.235701} {\bibfield
  {journal} {\bibinfo  {journal} {Phys. Rev. Lett}\ }\textbf {\bibinfo {volume}
  {93}},\ \bibinfo {pages} {235701:1} (\bibinfo {year} {2004})}\BibitemShut
  {NoStop}%
\bibitem [{\citenamefont {van Beest}\ \emph {et~al.}(1990)\citenamefont {van
  Beest}, \citenamefont {Kramer},\ and\ \citenamefont {van
  Santen}}]{BeestPhysRevLett1955}%
  \BibitemOpen
  \bibfield  {author} {\bibinfo {author} {\bibfnamefont {B.}~\bibnamefont {van
  Beest}}, \bibinfo {author} {\bibfnamefont {G.}~\bibnamefont {Kramer}},\ and\
  \bibinfo {author} {\bibfnamefont {R.}~\bibnamefont {van Santen}},\
  }\href@noop {} {\bibfield  {journal} {\bibinfo  {journal} {Phys. Rev. Lett.}\
  }\textbf {\bibinfo {volume} {64}},\ \bibinfo {pages} {1955} (\bibinfo {year}
  {1990})}\BibitemShut {NoStop}%
\bibitem [{\citenamefont {Yu}\ \emph {et~al.}(2021)\citenamefont {Yu},
  \citenamefont {Liu}, \citenamefont {Szlufarska},\ and\ \citenamefont
  {Wang}}]{Yu_Wang_PhysRevMater_2021}%
  \BibitemOpen
  \bibfield  {author} {\bibinfo {author} {\bibfnamefont {Z.}~\bibnamefont
  {Yu}}, \bibinfo {author} {\bibfnamefont {Q.}~\bibnamefont {Liu}}, \bibinfo
  {author} {\bibfnamefont {I.}~\bibnamefont {Szlufarska}},\ and\ \bibinfo
  {author} {\bibfnamefont {B.}~\bibnamefont {Wang}},\ }\href
  {https://doi.org/10.1103/PhysRevMaterials.5.015602} {\bibfield  {journal}
  {\bibinfo  {journal} {Phys. Rev. Materials}\ }\textbf {\bibinfo {volume}
  {5}},\ \bibinfo {pages} {015602} (\bibinfo {year} {2021})}\BibitemShut
  {NoStop}%
\bibitem [{\citenamefont {Yu}\ \emph {et~al.}(2022)\citenamefont {Yu},
  \citenamefont {Morgan}, \citenamefont {Ediger},\ and\ \citenamefont
  {Wang}}]{Yu_Wang_arxiv_2022}%
  \BibitemOpen
  \bibfield  {author} {\bibinfo {author} {\bibfnamefont {Z.}~\bibnamefont
  {Yu}}, \bibinfo {author} {\bibfnamefont {D.}~\bibnamefont {Morgan}}, \bibinfo
  {author} {\bibfnamefont {M.~D.}\ \bibnamefont {Ediger}},\ and\ \bibinfo
  {author} {\bibfnamefont {B.}~\bibnamefont {Wang}},\ }\href@noop {} {\bibfield
   {journal} {\bibinfo  {journal} {arXiv:2202.11000v1}\ } (\bibinfo {year}
  {2022})}\BibitemShut {NoStop}%
\bibitem [{\citenamefont {Tanaka}(2003)}]{Tanaka_JPhysCondMatt_2003}%
  \BibitemOpen
  \bibfield  {author} {\bibinfo {author} {\bibfnamefont {H.}~\bibnamefont
  {Tanaka}},\ }\href {https://doi.org/10.1088/0953-8984/15/45/l03} {\bibfield
  {journal} {\bibinfo  {journal} {Journal of Physics: Condensed Matter}\
  }\textbf {\bibinfo {volume} {15}},\ \bibinfo {pages} {L703} (\bibinfo {year}
  {2003})}\BibitemShut {NoStop}%
\bibitem [{\citenamefont {Shi}\ \emph {et~al.}(2018{\natexlab{a}})\citenamefont
  {Shi}, \citenamefont {Russo},\ and\ \citenamefont
  {Tanaka}}]{Shi_Tanaka_PNAS_2018}%
  \BibitemOpen
  \bibfield  {author} {\bibinfo {author} {\bibfnamefont {R.}~\bibnamefont
  {Shi}}, \bibinfo {author} {\bibfnamefont {J.}~\bibnamefont {Russo}},\ and\
  \bibinfo {author} {\bibfnamefont {H.}~\bibnamefont {Tanaka}},\ }\href
  {https://doi.org/10.1073/pnas.1807821115} {\bibfield  {journal} {\bibinfo
  {journal} {Proceedings of the National Academy of Sciences}\ }\textbf
  {\bibinfo {volume} {115}},\ \bibinfo {pages} {9444} (\bibinfo {year}
  {2018}{\natexlab{a}})}\BibitemShut {NoStop}%
\bibitem [{\citenamefont {Shi}\ \emph {et~al.}(2018{\natexlab{b}})\citenamefont
  {Shi}, \citenamefont {Russo},\ and\ \citenamefont
  {Tanaka}}]{Shi_Tanaka_JChemPhys_2018}%
  \BibitemOpen
  \bibfield  {author} {\bibinfo {author} {\bibfnamefont {R.}~\bibnamefont
  {Shi}}, \bibinfo {author} {\bibfnamefont {J.}~\bibnamefont {Russo}},\ and\
  \bibinfo {author} {\bibfnamefont {H.}~\bibnamefont {Tanaka}},\ }\href
  {https://doi.org/10.1063/1.5055908} {\bibfield  {journal} {\bibinfo
  {journal} {The Journal of Chemical Physics}\ }\textbf {\bibinfo {volume}
  {149}},\ \bibinfo {pages} {224502} (\bibinfo {year}
  {2018}{\natexlab{b}})}\BibitemShut {NoStop}%
\bibitem [{\citenamefont {Zeidler}\ \emph {et~al.}(2015)\citenamefont
  {Zeidler}, \citenamefont {Chirawatkul}, \citenamefont {Salmon}, \citenamefont
  {Usuki}, \citenamefont {Kohara}, \citenamefont {Fischer},\ and\ \citenamefont
  {Howells}}]{Zeidler_Howells_JNonCrystSolids_2015}%
  \BibitemOpen
  \bibfield  {author} {\bibinfo {author} {\bibfnamefont {A.}~\bibnamefont
  {Zeidler}}, \bibinfo {author} {\bibfnamefont {P.}~\bibnamefont
  {Chirawatkul}}, \bibinfo {author} {\bibfnamefont {P.~S.}\ \bibnamefont
  {Salmon}}, \bibinfo {author} {\bibfnamefont {T.}~\bibnamefont {Usuki}},
  \bibinfo {author} {\bibfnamefont {S.}~\bibnamefont {Kohara}}, \bibinfo
  {author} {\bibfnamefont {H.~E.}\ \bibnamefont {Fischer}},\ and\ \bibinfo
  {author} {\bibfnamefont {W.~S.}\ \bibnamefont {Howells}},\ }\href
  {https://doi.org/https://doi.org/10.1016/j.jnoncrysol.2014.08.027} {\bibfield
   {journal} {\bibinfo  {journal} {Journal of Non-Crystalline Solids}\ }\textbf
  {\bibinfo {volume} {407}},\ \bibinfo {pages} {235} (\bibinfo {year}
  {2015})},\ \bibinfo {note} {7th IDMRCS: Relaxation in Complex
  Systems}\BibitemShut {NoStop}%
\bibitem [{\citenamefont {Petri}\ \emph {et~al.}(2000)\citenamefont {Petri},
  \citenamefont {Salmon},\ and\ \citenamefont
  {Fischer}}]{Petri_Fischer_PhysRevLett_2000}%
  \BibitemOpen
  \bibfield  {author} {\bibinfo {author} {\bibfnamefont {I.}~\bibnamefont
  {Petri}}, \bibinfo {author} {\bibfnamefont {P.~S.}\ \bibnamefont {Salmon}},\
  and\ \bibinfo {author} {\bibfnamefont {H.~E.}\ \bibnamefont {Fischer}},\
  }\href {https://doi.org/10.1103/PhysRevLett.84.2413} {\bibfield  {journal}
  {\bibinfo  {journal} {Phys. Rev. Lett.}\ }\textbf {\bibinfo {volume} {84}},\
  \bibinfo {pages} {2413} (\bibinfo {year} {2000})}\BibitemShut {NoStop}%
\bibitem [{\citenamefont {Crichton}\ \emph {et~al.}(2001)\citenamefont
  {Crichton}, \citenamefont {Mezouar}, \citenamefont {Grande}, \citenamefont
  {St{\o}len},\ and\ \citenamefont
  {Grzechnik}}]{Crichton_Grzechnik_Nature_2001}%
  \BibitemOpen
  \bibfield  {author} {\bibinfo {author} {\bibfnamefont {W.~A.}\ \bibnamefont
  {Crichton}}, \bibinfo {author} {\bibfnamefont {M.}~\bibnamefont {Mezouar}},
  \bibinfo {author} {\bibfnamefont {T.}~\bibnamefont {Grande}}, \bibinfo
  {author} {\bibfnamefont {S.}~\bibnamefont {St{\o}len}},\ and\ \bibinfo
  {author} {\bibfnamefont {A.}~\bibnamefont {Grzechnik}},\ }\href
  {https://doi.org/10.1038/414622a} {\bibfield  {journal} {\bibinfo  {journal}
  {Nature}\ }\textbf {\bibinfo {volume} {414}},\ \bibinfo {pages} {622 }
  (\bibinfo {year} {2001})}\BibitemShut {NoStop}%
\bibitem [{\citenamefont {Edwards}\ and\ \citenamefont
  {Sen}(2011)}]{Edwards_Sen_JPhysChemB_2011}%
  \BibitemOpen
  \bibfield  {author} {\bibinfo {author} {\bibfnamefont {T.~G.}\ \bibnamefont
  {Edwards}}\ and\ \bibinfo {author} {\bibfnamefont {S.}~\bibnamefont {Sen}},\
  }\href {https://doi.org/10.1021/jp202174x} {\bibfield  {journal} {\bibinfo
  {journal} {The Journal of Physical Chemistry B}\ }\textbf {\bibinfo {volume}
  {115}},\ \bibinfo {pages} {4307} (\bibinfo {year} {2011})},\ \bibinfo {note}
  {pMID: 21446741}\BibitemShut {NoStop}%
\bibitem [{\citenamefont {Gupta}\ and\ \citenamefont
  {Mauro}(2009)}]{Gupta_Mauro_JChemPhys_2009}%
  \BibitemOpen
  \bibfield  {author} {\bibinfo {author} {\bibfnamefont {P.~K.}\ \bibnamefont
  {Gupta}}\ and\ \bibinfo {author} {\bibfnamefont {J.~C.}\ \bibnamefont
  {Mauro}},\ }\href {https://doi.org/10.1063/1.3077168} {\bibfield  {journal}
  {\bibinfo  {journal} {The Journal of Chemical Physics}\ }\textbf {\bibinfo
  {volume} {130}},\ \bibinfo {pages} {094503} (\bibinfo {year}
  {2009})}\BibitemShut {NoStop}%
\bibitem [{\citenamefont {Wilson}\ and\ \citenamefont
  {Salmon}(2009)}]{Wilson_Salmon_PhysRevLett_2009}%
  \BibitemOpen
  \bibfield  {author} {\bibinfo {author} {\bibfnamefont {M.}~\bibnamefont
  {Wilson}}\ and\ \bibinfo {author} {\bibfnamefont {P.~S.}\ \bibnamefont
  {Salmon}},\ }\href {https://doi.org/10.1103/PhysRevLett.103.157801}
  {\bibfield  {journal} {\bibinfo  {journal} {Phys. Rev. Lett.}\ }\textbf
  {\bibinfo {volume} {103}},\ \bibinfo {pages} {157801} (\bibinfo {year}
  {2009})}\BibitemShut {NoStop}%
\bibitem [{\citenamefont {Rosa~Junior}\ \emph {et~al.}(2019)\citenamefont
  {Rosa~Junior}, \citenamefont {Cruz}, \citenamefont {Santana},\ and\
  \citenamefont {Moret}}]{Rosa_Moret_PhysRevE_2019}%
  \BibitemOpen
  \bibfield  {author} {\bibinfo {author} {\bibfnamefont {A.~C.~P.}\
  \bibnamefont {Rosa~Junior}}, \bibinfo {author} {\bibfnamefont
  {C.}~\bibnamefont {Cruz}}, \bibinfo {author} {\bibfnamefont {W.~S.}\
  \bibnamefont {Santana}},\ and\ \bibinfo {author} {\bibfnamefont {M.~A.}\
  \bibnamefont {Moret}},\ }\href {https://doi.org/10.1103/PhysRevE.100.022139}
  {\bibfield  {journal} {\bibinfo  {journal} {Phys. Rev. E}\ }\textbf {\bibinfo
  {volume} {100}},\ \bibinfo {pages} {022139} (\bibinfo {year}
  {2019})}\BibitemShut {NoStop}%
\bibitem [{\citenamefont {Rosa}\ \emph {et~al.}(2020)\citenamefont {Rosa},
  \citenamefont {Cruz}, \citenamefont {Santana}, \citenamefont {Brito},\ and\
  \citenamefont {Moret}}]{Rosa_Moret_PhysRevE_2020}%
  \BibitemOpen
  \bibfield  {author} {\bibinfo {author} {\bibfnamefont {A.~C.~P.}\
  \bibnamefont {Rosa}}, \bibinfo {author} {\bibfnamefont {C.}~\bibnamefont
  {Cruz}}, \bibinfo {author} {\bibfnamefont {W.~S.}\ \bibnamefont {Santana}},
  \bibinfo {author} {\bibfnamefont {E.}~\bibnamefont {Brito}},\ and\ \bibinfo
  {author} {\bibfnamefont {M.~A.}\ \bibnamefont {Moret}},\ }\href
  {https://doi.org/10.1103/PhysRevE.101.042131} {\bibfield  {journal} {\bibinfo
   {journal} {Phys. Rev. E}\ }\textbf {\bibinfo {volume} {101}},\ \bibinfo
  {pages} {042131} (\bibinfo {year} {2020})}\BibitemShut {NoStop}%
\bibitem [{\citenamefont {Shi}\ and\ \citenamefont
  {Tanaka}(2018{\natexlab{b}})}]{Shi_Tanaka_PNAS_2018_2}%
  \BibitemOpen
  \bibfield  {author} {\bibinfo {author} {\bibfnamefont {R.}~\bibnamefont
  {Shi}}\ and\ \bibinfo {author} {\bibfnamefont {H.}~\bibnamefont {Tanaka}},\
  }\href {https://doi.org/10.1073/pnas.1717233115} {\bibfield  {journal}
  {\bibinfo  {journal} {Proceedings of the National Academy of Sciences}\
  }\textbf {\bibinfo {volume} {115}},\ \bibinfo {pages} {1980} (\bibinfo {year}
  {2018}{\natexlab{b}})}\BibitemShut {NoStop}%
\bibitem [{\citenamefont {Saksaengwijit}\ and\ \citenamefont
  {Heuer}(2006)}]{Heuer_PhysRevE_2006}%
  \BibitemOpen
  \bibfield  {author} {\bibinfo {author} {\bibfnamefont {A.}~\bibnamefont
  {Saksaengwijit}}\ and\ \bibinfo {author} {\bibfnamefont {A.}~\bibnamefont
  {Heuer}},\ }\href {https://doi.org/10.1103/PhysRevE.73.061503} {\bibfield
  {journal} {\bibinfo  {journal} {Phys. Rev. E}\ }\textbf {\bibinfo {volume}
  {73}},\ \bibinfo {pages} {061503} (\bibinfo {year} {2006})}\BibitemShut
  {NoStop}%
\bibitem [{\citenamefont {Lichtenstein}\ \emph
  {et~al.}(2012{\natexlab{a}})\citenamefont {Lichtenstein}, \citenamefont
  {B\"uchner}, \citenamefont {Yang}, \citenamefont {Shaikhutdinov},
  \citenamefont {Heyde}, \citenamefont {Sierka}, \citenamefont {W\l{}odarczyk},
  \citenamefont {Sauer},\ and\ \citenamefont {Freund}}]{HeydeAngewChem2012}%
  \BibitemOpen
  \bibfield  {author} {\bibinfo {author} {\bibfnamefont {L.}~\bibnamefont
  {Lichtenstein}}, \bibinfo {author} {\bibfnamefont {C.}~\bibnamefont
  {B\"uchner}}, \bibinfo {author} {\bibfnamefont {B.}~\bibnamefont {Yang}},
  \bibinfo {author} {\bibfnamefont {S.}~\bibnamefont {Shaikhutdinov}}, \bibinfo
  {author} {\bibfnamefont {M.}~\bibnamefont {Heyde}}, \bibinfo {author}
  {\bibfnamefont {M.}~\bibnamefont {Sierka}}, \bibinfo {author} {\bibfnamefont
  {R.}~\bibnamefont {W\l{}odarczyk}}, \bibinfo {author} {\bibfnamefont
  {J.}~\bibnamefont {Sauer}},\ and\ \bibinfo {author} {\bibfnamefont {H.-J.}\
  \bibnamefont {Freund}},\ }\href {https://doi.org/10.1002/anie.201107097}
  {\bibfield  {journal} {\bibinfo  {journal} {J. Angew. Chem. Int. Ed.}\
  }\textbf {\bibinfo {volume} {51}},\ \bibinfo {pages} {404} (\bibinfo {year}
  {2012}{\natexlab{a}})}\BibitemShut {NoStop}%
\bibitem [{\citenamefont {Lichtenstein}\ \emph
  {et~al.}(2012{\natexlab{b}})\citenamefont {Lichtenstein}, \citenamefont
  {Heyde},\ and\ \citenamefont {Freund}}]{HeydeJPhysChemC2012}%
  \BibitemOpen
  \bibfield  {author} {\bibinfo {author} {\bibfnamefont {L.}~\bibnamefont
  {Lichtenstein}}, \bibinfo {author} {\bibfnamefont {M.}~\bibnamefont
  {Heyde}},\ and\ \bibinfo {author} {\bibfnamefont {H.-J.}\ \bibnamefont
  {Freund}},\ }\href {https://doi.org/10.1021/jp3062866} {\bibfield  {journal}
  {\bibinfo  {journal} {J. Phys. Chem. C}\ }\textbf {\bibinfo {volume} {116}},\
  \bibinfo {pages} {20426} (\bibinfo {year} {2012}{\natexlab{b}})}\BibitemShut
  {NoStop}%
\bibitem [{\citenamefont {Huang}\ \emph {et~al.}(2012)\citenamefont {Huang},
  \citenamefont {Kurasch}, \citenamefont {Srivastava}, \citenamefont
  {Skakalova}, \citenamefont {Kotakoski}, \citenamefont {Krasheninnikov},
  \citenamefont {Hovden}, \citenamefont {Mao}, \citenamefont {Meyer},
  \citenamefont {Smet}, \citenamefont {Muller},\ and\ \citenamefont
  {Kaiser}}]{Huang_Kaiser_NanoLett_2012}%
  \BibitemOpen
  \bibfield  {author} {\bibinfo {author} {\bibfnamefont {P.~Y.}\ \bibnamefont
  {Huang}}, \bibinfo {author} {\bibfnamefont {S.}~\bibnamefont {Kurasch}},
  \bibinfo {author} {\bibfnamefont {A.}~\bibnamefont {Srivastava}}, \bibinfo
  {author} {\bibfnamefont {V.}~\bibnamefont {Skakalova}}, \bibinfo {author}
  {\bibfnamefont {J.}~\bibnamefont {Kotakoski}}, \bibinfo {author}
  {\bibfnamefont {A.~V.}\ \bibnamefont {Krasheninnikov}}, \bibinfo {author}
  {\bibfnamefont {R.}~\bibnamefont {Hovden}}, \bibinfo {author} {\bibfnamefont
  {Q.}~\bibnamefont {Mao}}, \bibinfo {author} {\bibfnamefont {J.~C.}\
  \bibnamefont {Meyer}}, \bibinfo {author} {\bibfnamefont {J.}~\bibnamefont
  {Smet}}, \bibinfo {author} {\bibfnamefont {D.~A.}\ \bibnamefont {Muller}},\
  and\ \bibinfo {author} {\bibfnamefont {U.}~\bibnamefont {Kaiser}},\ }\href
  {https://doi.org/10.1021/nl204423x} {\bibfield  {journal} {\bibinfo
  {journal} {Nano Lett.}\ }\textbf {\bibinfo {volume} {12}},\ \bibinfo {pages}
  {1081} (\bibinfo {year} {2012})}\BibitemShut {NoStop}%
\bibitem [{\citenamefont {Heyde}\ \emph {et~al.}(2012)\citenamefont {Heyde},
  \citenamefont {Shaikhutdinov},\ and\ \citenamefont
  {Freund}}]{HeydeChemPhysLett2012}%
  \BibitemOpen
  \bibfield  {author} {\bibinfo {author} {\bibfnamefont {M.}~\bibnamefont
  {Heyde}}, \bibinfo {author} {\bibfnamefont {S.}~\bibnamefont
  {Shaikhutdinov}},\ and\ \bibinfo {author} {\bibfnamefont {H.-J.}\
  \bibnamefont {Freund}},\ }\href
  {https://doi.org/10.1016/j.cplett.2012.08.063} {\bibfield  {journal}
  {\bibinfo  {journal} {Chem. Phys. Lett.}\ }\textbf {\bibinfo {volume}
  {550}},\ \bibinfo {pages} {1} (\bibinfo {year} {2012})}\BibitemShut {NoStop}%
\bibitem [{\citenamefont {Roy}\ \emph {et~al.}(2018)\citenamefont {Roy},
  \citenamefont {Heyde},\ and\ \citenamefont {Heuer}}]{Roy_Heuer_PCCP_2018}%
  \BibitemOpen
  \bibfield  {author} {\bibinfo {author} {\bibfnamefont {P.~K.}\ \bibnamefont
  {Roy}}, \bibinfo {author} {\bibfnamefont {M.}~\bibnamefont {Heyde}},\ and\
  \bibinfo {author} {\bibfnamefont {A.}~\bibnamefont {Heuer}},\ }\href
  {https://doi.org/10.1039/C8CP01313F} {\bibfield  {journal} {\bibinfo
  {journal} {Phys. Chem. Chem. Phys.}\ }\textbf {\bibinfo {volume} {20}},\
  \bibinfo {pages} {14725} (\bibinfo {year} {2018})}\BibitemShut {NoStop}%
\bibitem [{\citenamefont {Roy}\ and\ \citenamefont
  {Heuer}(2019{\natexlab{a}})}]{Roy_Heuer_PRL_2019}%
  \BibitemOpen
  \bibfield  {author} {\bibinfo {author} {\bibfnamefont {P.~K.}\ \bibnamefont
  {Roy}}\ and\ \bibinfo {author} {\bibfnamefont {A.}~\bibnamefont {Heuer}},\
  }\href {https://doi.org/10.1103/PhysRevLett.122.016104} {\bibfield  {journal}
  {\bibinfo  {journal} {Phys. Rev. Lett.}\ }\textbf {\bibinfo {volume} {122}},\
  \bibinfo {pages} {016104} (\bibinfo {year} {2019}{\natexlab{a}})}\BibitemShut
  {NoStop}%
\bibitem [{\citenamefont {Roy}\ and\ \citenamefont
  {Heuer}(2019{\natexlab{b}})}]{Roy_Heuer_JPhysCondMatt_2019}%
  \BibitemOpen
  \bibfield  {author} {\bibinfo {author} {\bibfnamefont {P.~K.}\ \bibnamefont
  {Roy}}\ and\ \bibinfo {author} {\bibfnamefont {A.}~\bibnamefont {Heuer}},\
  }\href {https://doi.org/10.1088/1361-648x/ab0a13} {\bibfield  {journal}
  {\bibinfo  {journal} {Journal of Physics: Condensed Matter}\ }\textbf
  {\bibinfo {volume} {31}},\ \bibinfo {pages} {225703} (\bibinfo {year}
  {2019}{\natexlab{b}})}\BibitemShut {NoStop}%
\bibitem [{\citenamefont {Nos\'e}(1984)}]{NoseJChemPhys1984}%
  \BibitemOpen
  \bibfield  {author} {\bibinfo {author} {\bibfnamefont {S.}~\bibnamefont
  {Nos\'e}},\ }\href {https://doi.org/10.1063/1.447334} {\bibfield  {journal}
  {\bibinfo  {journal} {J, Chem. Phys.}\ }\textbf {\bibinfo {volume} {81}},\
  \bibinfo {pages} {511} (\bibinfo {year} {1984})}\BibitemShut {NoStop}%
\bibitem [{\citenamefont {Martyna}\ \emph {et~al.}(1992)\citenamefont
  {Martyna}, \citenamefont {Klein},\ and\ \citenamefont
  {Tuckerman}}]{MartynaJChemPhys1992}%
  \BibitemOpen
  \bibfield  {author} {\bibinfo {author} {\bibfnamefont {G.~J.}\ \bibnamefont
  {Martyna}}, \bibinfo {author} {\bibfnamefont {M.~L.}\ \bibnamefont {Klein}},\
  and\ \bibinfo {author} {\bibfnamefont {M.}~\bibnamefont {Tuckerman}},\ }\href
  {https://doi.org/10.1063/1.463940} {\bibfield  {journal} {\bibinfo  {journal}
  {The Journal of Chemical Physics}\ }\textbf {\bibinfo {volume} {97}},\
  \bibinfo {pages} {2635} (\bibinfo {year} {1992})}\BibitemShut {NoStop}%
\bibitem [{\citenamefont {B\"uchner}\ and\ \citenamefont
  {Heuer}(1999)}]{HeuerPhysRevE1999}%
  \BibitemOpen
  \bibfield  {author} {\bibinfo {author} {\bibfnamefont {S.}~\bibnamefont
  {B\"uchner}}\ and\ \bibinfo {author} {\bibfnamefont {A.}~\bibnamefont
  {Heuer}},\ }\href {https://doi.org/10.1103/PhysRevE.60.6507} {\bibfield
  {journal} {\bibinfo  {journal} {Phys. Rev. E}\ }\textbf {\bibinfo {volume}
  {60}},\ \bibinfo {pages} {6507} (\bibinfo {year} {1999})}\BibitemShut
  {NoStop}%
\bibitem [{\citenamefont {Heuer}\ and\ \citenamefont
  {BÃ¼chner}(2000)}]{HeuerJPhysCondMatt2000}%
  \BibitemOpen
  \bibfield  {author} {\bibinfo {author} {\bibfnamefont {A.}~\bibnamefont
  {Heuer}}\ and\ \bibinfo {author} {\bibfnamefont {S.}~\bibnamefont
  {BÃ¼chner}},\ }\href@noop {} {\bibfield  {journal} {\bibinfo  {journal}
  {Journal of Physics: Condensed Matter}\ }\textbf {\bibinfo {volume} {12}},\
  \bibinfo {pages} {6535} (\bibinfo {year} {2000})}\BibitemShut {NoStop}%
\bibitem [{\citenamefont {Heuer}\ and\ \citenamefont
  {Saksaengwijit}(2008)}]{HeuerPhysRevE2008}%
  \BibitemOpen
  \bibfield  {author} {\bibinfo {author} {\bibfnamefont {A.}~\bibnamefont
  {Heuer}}\ and\ \bibinfo {author} {\bibfnamefont {A.}~\bibnamefont
  {Saksaengwijit}},\ }\href {https://doi.org/10.1103/PhysRevE.77.061507}
  {\bibfield  {journal} {\bibinfo  {journal} {Phys. Rev. E}\ }\textbf {\bibinfo
  {volume} {77}},\ \bibinfo {pages} {061507} (\bibinfo {year}
  {2008})}\BibitemShut {NoStop}%
\bibitem [{\citenamefont {Kumar}\ \emph {et~al.}(1992)\citenamefont {Kumar},
  \citenamefont {Rosenberg}, \citenamefont {Bouzida}, \citenamefont
  {Swendsen},\ and\ \citenamefont {Kollman}}]{Kumar_Kollman_JCompChem_1992}%
  \BibitemOpen
  \bibfield  {author} {\bibinfo {author} {\bibfnamefont {S.}~\bibnamefont
  {Kumar}}, \bibinfo {author} {\bibfnamefont {J.~M.}\ \bibnamefont
  {Rosenberg}}, \bibinfo {author} {\bibfnamefont {D.}~\bibnamefont {Bouzida}},
  \bibinfo {author} {\bibfnamefont {R.~H.}\ \bibnamefont {Swendsen}},\ and\
  \bibinfo {author} {\bibfnamefont {P.~A.}\ \bibnamefont {Kollman}},\ }\href
  {https://doi.org/https://doi.org/10.1002/jcc.540130812} {\bibfield  {journal}
  {\bibinfo  {journal} {Journal of Computational Chemistry}\ }\textbf {\bibinfo
  {volume} {13}},\ \bibinfo {pages} {1011} (\bibinfo {year}
  {1992})}\BibitemShut {NoStop}%
\bibitem [{\citenamefont {Gallicchio}\ \emph {et~al.}(2005)\citenamefont
  {Gallicchio}, \citenamefont {Andrec}, \citenamefont {Felts},\ and\
  \citenamefont {Levy}}]{GallicchioJPhysChemB2005}%
  \BibitemOpen
  \bibfield  {author} {\bibinfo {author} {\bibfnamefont {E.}~\bibnamefont
  {Gallicchio}}, \bibinfo {author} {\bibfnamefont {M.}~\bibnamefont {Andrec}},
  \bibinfo {author} {\bibfnamefont {A.~K.}\ \bibnamefont {Felts}},\ and\
  \bibinfo {author} {\bibfnamefont {R.~M.}\ \bibnamefont {Levy}},\ }\href
  {https://doi.org/10.1021/jp045294f} {\bibfield  {journal} {\bibinfo
  {journal} {J. Phys. Chem. B}\ }\textbf {\bibinfo {volume} {109}},\ \bibinfo
  {pages} {6722} (\bibinfo {year} {2005})}\BibitemShut {NoStop}%
\bibitem [{\citenamefont {Doliwa}\ and\ \citenamefont
  {Heuer}(2003{\natexlab{a}})}]{DoliwaPhysRevLett2003}%
  \BibitemOpen
  \bibfield  {author} {\bibinfo {author} {\bibfnamefont {B.}~\bibnamefont
  {Doliwa}}\ and\ \bibinfo {author} {\bibfnamefont {A.}~\bibnamefont {Heuer}},\
  }\href {https://doi.org/10.1103/PhysRevLett.91.235501} {\bibfield  {journal}
  {\bibinfo  {journal} {Phys. Rev. Lett.}\ }\textbf {\bibinfo {volume} {91}},\
  \bibinfo {pages} {235501} (\bibinfo {year} {2003}{\natexlab{a}})}\BibitemShut
  {NoStop}%
\bibitem [{\citenamefont {Doliwa}\ and\ \citenamefont
  {Heuer}(2003{\natexlab{b}})}]{DoliwaPhysRevE2003}%
  \BibitemOpen
  \bibfield  {author} {\bibinfo {author} {\bibfnamefont {B.}~\bibnamefont
  {Doliwa}}\ and\ \bibinfo {author} {\bibfnamefont {A.}~\bibnamefont {Heuer}},\
  }\href {https://doi.org/10.1103/PhysRevE.67.031506} {\bibfield  {journal}
  {\bibinfo  {journal} {Phys. Rev. E}\ }\textbf {\bibinfo {volume} {67}},\
  \bibinfo {pages} {031506} (\bibinfo {year} {2003}{\natexlab{b}})}\BibitemShut
  {NoStop}%
\bibitem [{\citenamefont {Stillinger}\ and\ \citenamefont
  {Weber}(1985)}]{StillingerPhysRevB1985}%
  \BibitemOpen
  \bibfield  {author} {\bibinfo {author} {\bibfnamefont {F.~H.}\ \bibnamefont
  {Stillinger}}\ and\ \bibinfo {author} {\bibfnamefont {T.~A.}\ \bibnamefont
  {Weber}},\ }\href {https://doi.org/10.1103/PhysRevB.31.5262} {\bibfield
  {journal} {\bibinfo  {journal} {Phys Rev B}\ }\textbf {\bibinfo {volume}
  {31}},\ \bibinfo {pages} {5262} (\bibinfo {year} {1985})}\BibitemShut
  {NoStop}%
\end{thebibliography}

\end{document}